\documentclass[a4paper,fleqn,usenatbib]{mnras}

\usepackage{newtxtext}
\usepackage{newtxmath}

\usepackage[T1]{fontenc}
\usepackage{ae,aecompl}

\usepackage{booktabs}

\newcommand{\Jupiter}{\textsc{j}} %
\newcommand{\Planet}{\textsc{p}} %
\newcommand{\Sun}{{\mathchoice{}{}{\scriptscriptstyle}{}\odot}} %
\newcommand{\Star}{{\mathchoice{}{}{\scriptscriptstyle}{}\star}} %
\newcommand{\Alfven}{\textsc{a}} %
\newcommand{\Aurora}{{\text{a}}}  %
\newcommand{\Wind}{\textsc{w}} %
\newcommand{\Mag}{\text{m}} %
\newcommand{\Magnetic}{{\scriptscriptstyle{\!B}}} %
\newcommand{\Thermal}{\text{th}} %
\newcommand{\Ram}{\text{ram}} %
\newcommand{\Fast}{\text{fast}} %
\newcommand{\Kinetic}{\text{k}} %
\newcommand{\Gyro}{\text{ce}}  %
\newcommand{\Plasma}{\text{pe}}  %
\newcommand{\Electron}{\text{e}}  %
\newcommand{\awsom}{AWSoM}
\newcommand{\swmf}{SWMF}
\newcommand{\batsrus}{BATS-R-US}

\usepackage{orcidlink} %

\usepackage{siunitx}
\DeclareSIUnit \parsec{pc}
\DeclareSIUnit \erg{erg}
\DeclareSIUnit \gauss{G}
\DeclareSIUnit \Rsun{R_\Sun}
\DeclareSIUnit \Msun{M_\Sun}
\DeclareSIUnit \Rstar{R_\Star}
\DeclareSIUnit \Rjup{R_\Jupiter}
\DeclareSIUnit \Mjup{M_\Jupiter}
\DeclareSIUnit{\jansky}{Jy}

\usepackage{color, soul, ulem,graphicx, amsbsy, xcolor}

\renewcommand{\vec}[1]{\boldsymbol{#1}} %
\newcommand{\uvec}[1]{\boldsymbol{\hat{#1}}} %

\title[Contemporaneous wind model of radio emission from $\tau$~Boötis]{An observationally based wind model contemporaneous with the radio detections in \(\tau\)~Boötis.}

\author[D. Evensberget et al.]{
    D. Evensberget~\orcidlink{0000-0001-7810-8028}\(^{1}\)\thanks{E-mail: \href{mailto:evensberget@strw.leidenuniv.nl}{evensberget@strw.leidenuniv.nl}},
    A. A. Vidotto~\orcidlink{0000-0001-5371-2675}\(^{1}\),
    F. Elekes~\orcidlink{0000-0002-7258-3386}\(^{1,2}\),
    S. V. Jeffers~\orcidlink{0000-0003-2490-4779}\(^{3}\)
    and
    R. T. Luisman\(^{1}\)
    \\
    \(^{1}\)Leiden Observatory, Leiden University, PO Box 9513, 2300 RA Leiden, The Netherlands
    \\
    \(^{2}\)Institute of Geophysics and Meteorology, University of Cologne, Pohligstr. 3, 50969 Köln, Germany
    \\
    \(^{3}\)Thüringer Landessternwarte Tautenburg, Sternwarte 5, 07778 Tautenburg, Germany
}

\date{Accepted XXX. Received YYY; in original form ZZZ}

\pubyear{}

\begin{document}
\label{firstpage}
\pagerange{\pageref{firstpage}--\pageref{lastpage}}
\maketitle

\begin{abstract}
Recent low-frequency array (LOFAR) radio signal detections bearing from the \(\tau\)~Boötis system have been cautiously attributed to auroral emissions from the hot Jupiter $\tau$~Boötis~Ab. The auroral emissions are believed to be excited by interaction between the exoplanet and the winds of its host star.
Since stellar winds respond to stellar surface magnetism, three-dimensional stellar wind modelling, able to account for the star's contemporaneous magnetic field geometry, can aid the interpretation of radio detections.
For the first time, we present spectropolarimetric observations of \(\tau\)~Boötis~A from the same epoch as the LOFAR detections. We derive a contemporaneous large-scale magnetic map of \(\tau\)~Boötis~A, which shows a poloidally dominated field with mean strength 1.6\,G.
From our magnetic map, we create a three-dimensional numerical wind model and characterise the wind properties around \(\tau\)~Boötis~Ab.
To compute the wind power dissipated in \(\tau\)~Boötis~Ab's magnetosphere, we apply two approaches:
A)~the solar system-based empirical relation called Bode's law; and
B)~a resolved numerical model of the planetary magnetosphere.
When consistently applying best-case assumptions, we predict radio flux densities around 50\,mJy and 0.68\,mJy respectively. Our values are much too small to be consistent with the reported observation of $890^{+690}_{-500}$\,mJy; a stellar surface magnetic field scaling \({\gtrsim}10\) is required to reproduce the observed signal strength.
As \(\tau\)~Boötis~A has a rapid magnetic cycle, we speculate that wind variations cased by variation in stellar magnetism may explain the lack of detections from follow-up observations. Our work emphasises the importance of contemporaneous observations of stellar magnetism and observational signatures of star-planet interaction.

\end{abstract}
\begin{keywords}
stars: planetary systems -- stars: low-mass -- stars: winds, outflows -- planet-star interactions
\end{keywords}

\section{Introduction}\label{sec.intro}

Recently, \citet{2021A&A...645A..59T} detected low-frequency radio signals bearing from the \(\tau\)~Boötis planetary system. The authors cautiously suggest that the emission originates from the hot~Jupiter \(\tau\)~Boötis~Ab, which has previously been identified as a favourable candidate for radio emissions~\citep{2005A&A...437..717G,2018MNRAS.480.3680W}. The signal detections, which were in the \qtyrange{10}{30}{\mega\hertz} frequency range, were conducted with the low-frequency array~\citep[LOFAR,][]{2013A&A...556A...2V}.

A plausible mechanism behind the radio emission is interaction between the wind of the host star \(\tau\)~Boötis~A and the planetary magnetosphere, resulting in auroral radio emissions from \(\tau\)~Boötis~Ab. In the auroral emission mechanism, energetic electrons are thought to emit radio waves via the electron cyclotron-maser instability~\citep[ECMI,][]{1958AuJPh..11..564T,1975JGR....80.4675S,1979ApJ...230..621W,2006A&ARv..13..229T,2015Natur.523..568H}. Such
ECMI emissions have been observed from the auroral regions of solar system planets~\citep{1958JGR....63..807F,1975JGR....80.4675S}.
The `radiometric Bode's law'~\citep{1984Natur.310..755D}
is a power law-type relation between observed radio emissions of the magnetised planets in the solar system and the incident power of the solar wind, which extends over several orders of magnitude~\citep[see][]{2001Ap&SS.277..293Z, 2018A&A...618A..84Z}. %

It is natural to ask whether magnetised exoplanets emit at radio frequencies in the same way that solar system planets do~\citep[see e.g.][]{1999JGR...10414025F,2000ApJ...545.1058B,2001Ap&SS.277..293Z,2004ApJ...612..511L}.
While a full treatment of the ECMI beaming pattern is complex~\citep{2011A&A...531A..29H,2019A&A...627A..30L,2023MNRAS.524.6267K}, as is a detailed treatment of the emission mechanism
itself~\citep{1984JGR....89.2831L,2010GeoRL..3719105M,2017GeoRL..44.4439L,2023pre9.conf03095C},
the radiometric Bode's law suggests that the emissions of hot Jupiter-type planets can be several orders of magnitude more powerful than those observed in the solar system~\citep[][]{2005A&A...437..717G, 2007A&A...475..359G,2007P&SS...55..598Z,2010ApJ...720.1262V, 2012MNRAS.423.3285V}.
If the radiometric Bode's law and similar scaling relations apply to exoplanetary systems, then the kinetic and magnetic energy fluxes derived from stellar wind modelling can inform the search for exoplanetary radio signals~\citep{2005A&A...437..717G,2017A&A...602A..39V,2019MNRAS.485.4529K}.The wind energy fluxes then permit us to estimate the total intensity of the emission, and simple geometrical modelling also permits us to roughly estimate the flux densities that would be observed by radio telescopes~\citep[following e.g.][]{2001Ap&SS.277..293Z,2017A&A...602A..39V}.

Radio detection of exoplanets holds promise as a new method of exoplanet detection~\citep[e.g.][]{2024arXiv240915507C}.
Furthermore, radio emissions from planetary magnetospheres provide a direct measure of the existence and strength of planetary magnetic fields which play an important role in planetary energy budgets and habitability~\citep[e.g.][]{2007AsBio...7..185L, 2019MNRAS.485.3999M}. In the future, radio detections may even be able to constrain the geometry of the planetary magnetic field~\citep[e.g.][]{2011A&A...531A..29H}. The strength of radio emissions would also provide information on the energy density of the planet's space environment, which could be used to further constrain stellar wind properties.

Stellar winds are shaped by the stellar surface magnetic field~\citep{2009ApJ...699..441V,2010ApJ...721...80C}, and \(\tau\)~Boötis~A is known for its variable magnetic field~\citep{2007MNRAS.374L..42C,2008MNRAS.385.1179D,2009MNRAS.398.1383F,2013MNRAS.435.1451F,2016MNRAS.459.4325M,2018MNRAS.479.5266J}.
We therefore expect the radio flux density from \(\tau\)~Boötis~Ab to vary depending on both the varying magnetic geometry and the planetary phase. We also expect the radio flux density to be affected by transient events such as coronal mass ejections and corotational breakdown.

In this work, we present for the first time a magnetic map of the host star \(\tau\)~Boötis~A that is contemporaneous with the radio signal detections reported by~\citet{2021A&A...645A..59T}.
By letting the contemporaneous magnetic map drive an Alfvén wave powered  numerical wind model~\citep{2013ApJ...764...23S,2014ApJ...782...81V}, we are able to model the space weather environment of the \(\tau\)~Boötis~A system from the stellar chromosphere and past the orbit of \(\tau\)~Boötis~Ab at the time of the LOFAR observations.
Our model lets us assess whether the planet is in the subalfvénic or superalfvénic regions of the stellar wind. With our stellar wind solution we assess the total energy flux of the wind at the position of \(\tau\)~Boötis~Ab incident on its magnetosphere.

Particulars of our solar system leads to a phenomenon where the giant planets face a wind whose kinetic energy is \qty{\sim 170} times greater than the wind magnetic energy; this fixed ratio in the outer solar system complicated the determination of the primary power source of radio emissions
\citep{2001Ap&SS.277..293Z,2007P&SS...55..598Z}. \citet{2018A&A...618A..84Z} showed, however, that the wind magnetic energy flux is sufficient to explain radio emissions from both, Solar system planets and the Jovian moons uniformly.

A solar system-like, fixed ratio between kinetic and magnetic energy flux is not expected in compact exoplanetary systems~\citep[see][]{2010ApJ...720.1262V}, and in this work we model both the kinetic and magnetic energy fluxes at the position of \(\tau\)~Boötis~Ab for the sake of better understanding the energetics in this different scenario.
We use the stellar wind properties at the location of \(\tau\)~Boötis~Ab to estimate the power of auroral radio emissions via two different approaches:
(A) By applying an unresolved Bode's law-based approach, driven by both the wind kinetic energy flux and Poynting flux~\citep[as in][]{2018A&A...618A..84Z}; and
(B) By driving a resolved planet-centred magnetohydrodynamical model of the wind-planet interaction as in the previous work on
    \(\tau\)~Boötis~Ab~\citep[see][]{2023A&A...671A.133E}.

From the auroral radio power we estimate the resulting signal flux density as seen by LOFAR.
This paper is organised as follows:
In Section~\ref{sec:tauBoo} we present the \(\tau\)~Boötis system and the radio observations of~\citet{2021A&A...645A..59T};
In Section~\ref{sec:spectropolarimetry} we present the contemporaneous spectropolarimetric observations and the resulting radial magnetic field map of \(\tau\)~Boötis~A;
In Section~\ref{sec:stellar-wind-models} we present the numerical wind models and we compute relevant wind properties at the planetary location;
In Section~\ref{sec:radio-emissions-power} we compute the available radio emissions power at the planetary location;
In Section~\ref{sec:radio-flux-densities} we compute the expected radio flux densities observed by LOFAR;
In Section~\ref{sec:discussion} we discuss the implications of our results; and
in Section~\ref{sec:conclusion} we present our conclusions.

\section[The tau Boötis system]{The \(\tau\)~Boötis system}\label{sec:tauBoo}
The \(\tau\)~Boötis system comprises a binary system of an F6V~dwarf \(\tau\)~Boötis~A
and an M2~dwarf \(\tau\)~Boötis~B  on a wide orbit~\citep{2019A&A...625A..59J}. The system is located at a distance of \qty{15.66\pm0.08}{\parsec} from the Sun. The primary \(\tau\)~Boötis~A has a detected planet, \(\tau\)~Boötis~Ab~\citep{1997ApJ...474L.115B}, which has a semi-major axis of about \(7.2 R_\Star\) (see Table~\ref{tab:properties}).  \(\tau\)~Boötis~Ab is thought to be tidally locked to \(\tau\)~Boötis~A, always presenting the same side towards the star. Furthermore, due to the similarity of the stellar rate of rotation with the orbital period of the planet, the planet is thought to be in synchronous rotation (maintained by tidal interactions) with the star~\citep{2005ApJ...622.1075S,2008ApJ...676..628S} so that the star also always presents the same side towards the planet.
\citet{2016MNRAS.459.4325M} found, based on spectropolarimetric observations, an equatorial rotation period of \qty{3.2\pm0.14}{\day}, and surface
differential rotation of up to \qty{20}{\percent} was found. The stellar rotation rate at \qty{40}{\degree} is expected to be equal to the orbital period of the planet~\citep{2009MNRAS.398.1383F,2016MNRAS.459.4325M,2018MNRAS.479.5266J}. These properties are summarised in the top half of Table~\ref{tab:properties}.

\begin{table}
    \caption{
        Properties of the \(\tau\)~Boötis system. The numbers in the `ref.' column refer to the following works:
        1: \citet{2021ApJS..255....8R},
        2: \citet{2007MNRAS.374L..42C},
        3: \citet{2016MNRAS.459.4325M},
        4: \citet{2023A&A...674A...1G}.
        5: \citet{2011MNRAS.418.1822W},
        6: \citet{2012Natur.486..502B},
        7: \citet{2015A&A...578A..64B},
        and
        8: \citet{2021A&A...645A..59T}.
    }\label{tab:properties}
    \centering
    \begin{tabular}{lrlr}
    \toprule
    Star \(\tau\)~Boötis~A  && Value & Ref. \\
    \midrule
    Radius                       & $R_\Star$ & \qty{1.4397\pm 0.0305}{\Rsun}        & 1 \\
    Mass                         & $M_\Star$ & \qty{1.4032\pm 0.0340}{\Msun}        & 1 \\
    Spin inclination             & $i_\Star$ & \qty{40}{\degree}                    & 2 \\
    Equatorial rotation period   & $P_\Star$ & \qty{3.2 \pm 0.14}{\day}             & 3 \\
    Distance from Earth          & $d_\Star$ & \qty{15.66\pm 0.08}{\parsec}         & 4 \\
    Magnetic field strength      & $B_\Star$ & \qty{1.21}{\gauss}                   & Fig.~\ref{fig:map} \\
    \midrule
    Planet \(\tau\)~Boötis~Ab  && Value & Ref. \\
    \midrule
    Orbital period           & $P_\Planet$      & \qty{3.312433(19)}{\day}  & 4 \\
    Epoch of periastron     & $T_\Planet$      & \qty{2456400.94\pm0.30}{\day}  & 5 \\
    Semi-major axis          & $a_\Planet$      & \qty{0.04869\pm 0.00039}{\astronomicalunit} & 1 \\
    Radius                   & $R_\Planet$      & \qty{1.06(1:2)}{\Rjup}                      & 5 \\
    Mass                     & $M_\Planet$      & \qty{4.300\pm0.075}{\Mjup}                  & 1 \\
    Eccentricity             & $e_\Planet$      & \qty{0.00740(0.00590:0.00480)}{\degree}     & 1 \\
    Inclination              & $i_\Planet$      & \qty{44.5 \pm 1.5}{\degree}                 & 6 \\
    Argument of periastron   & $\omega_\Planet$ & \qty{113.4\pm32.2}{\degree}                 & 7 \\
    Time of conjunction      & $T_\text{conj}$  & \qty{2455652.108\pm0.004}{\day}                   & 4 \\
    Magnetic field strength (max) & $B_\Planet$      & \qty{8\pm3}{\gauss}                         & 8 \\
    \bottomrule
    \end{tabular}
    \end{table}

Besides being a hot Jupiter host, the star \(\tau\)~Boötis~A is well known for its magnetic polarity reversals~\citep{2007MNRAS.374L..42C,2008MNRAS.385.1179D,2009MNRAS.398.1383F,2013MNRAS.435.1451F} suggesting a rapid magnetic cycle of \qty{\sim240}{\day}~\citep{2016MNRAS.459.4325M,2018MNRAS.479.5266J}.
The stellar winds of \(\tau\)~Boötis~A have been modelled numerically using a polytropic wind with a hot corona~\citep{2012MNRAS.423.3285V,2016MNRAS.459.1907N}
who found wind mass loss rates of
\qtyrange{2.3e-12}{2.7e-12}{\Msun/yr} varying through the magnetic cycle.
\citet{2012MNRAS.423.3285V} also applied a semi-analytical model of planetary auroral radio emissions and showed that \(\tau\)~Boötis~Ab could be a strong radio source observable by radio telescopes.

\(\tau\)~Boötis~Ab is one of the first discovered exoplanets~\citep{1997ApJ...474L.115B}, with a radius of \(R_\Planet=\qty{1.06(1:2)}{\Rjup}\)~\citep{2011MNRAS.418.1822W}, orbiting at a close-in distance of \qty{0.049}{\astronomicalunit} and period \qty{3.3}{\day}~\citep{2021ApJS..255....8R}. We adopt the orbital inclination \qty{44.5 \pm 1.5}{\degree} estimated by~\citet{2012Natur.486..502B}, noting that~\citet{2012ApJ...753L..25R} found~\qty{47(7:6)}{\degree} and that~\citet{2014ApJ...783L..29L} found~\qty{45(3:4)}{\degree} for this parameter.
The dayside surface temperature of \(\tau\)~Boötis~Ab is estimated to be \qty{1600}{\kelvin}~\citep{2008SPIE.7013E..2ZR}, but hydrodynamic escape is not expected due to the high mass of the planet and the resulting surface gravity~\citep{2018MNRAS.480.3680W}.
The planetary parameters adopted in this study are given in the bottom part of Table~\ref{tab:properties}.

The radio observations we consider in this work comprise seven \qtyrange{2.5}{3}{\hour} observations of the \(\tau\)~Boötis system made between 18 February 2017 and 25 March 2017~\citep[Appendix A in][]{2021A&A...645A..59T}. Two of these observations, L569131 (2017-02-18) and L570725 (2017-03-06), are tentative detections ascribed to \(\tau\)~Boötis~Ab. Of the two observations, L569131 is referred to as a `bursty' signal reaching an observed flux density of
\qty{890(690:500)}{\milli\jansky} in the \qtyrange{15}{21}{\mega\hertz} range, while L570725 is a `slowly varying' emission in the range \qtyrange{21}{30}{\mega\hertz} range. This signal has an average flux density of \qty{190\pm30}{\milli\jansky} and a peak flux of \qty{430\pm30}{\milli\jansky}. The slowly varying signal lasted for more than \qty{1}{\hour}.

Based on the frequency ranges in the detected radio signals, \citet{2021A&A...645A..59T} estimated the polar surface magnetic field strength of \(\tau\)~Boötis~Ab to be in the range from \qtyrange{5}{11}{\gauss}. For comparison, Jupiter's polar surface magnetic field strength is \qty{\sim14}{\gauss}~\citep{1976JGR....81.2917A,1993JGR....9818659C}. We note that modelling indicates that \(\tau\)~Boötis~Ab may have a stronger field than Jupiter, as the planetary magnetic field strength is affected by the stellar energy flux~\citep{2009Natur.457..167C,2024MNRAS.535.3646K}. In this work we apply a polar field strength of \qty{9}{G} for consistency with the modelling in~\citet{2023A&A...671A.133E}.
\section{Contemporaneous spectropolarimetric observations}\label{sec:spectropolarimetry}

\begin{figure}
    \centering
    \includegraphics{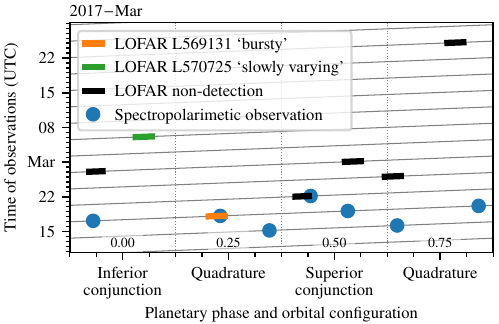}
    \caption{
        Time and orbital phase of the LOFAR observations of~\citet{2021A&A...645A..59T} and the spectropolarimetric observations presented in this work. For each observation, the orbital configuration can be read of the \(x\)-axis and the date can be read of the \(y\)-axis.
        The orbital configuration is based on the model of~\citet{2012Natur.486..502B}.
        The planetary configuration is shown as a grey line. The phase is zero at inferior conjunction, which is when \(\tau\)~Boötis~Ab is at its closest distance to the observer.
        The times of spectropolarimetric observations are indicated with blue dots (these times are also given in Table~\ref{tab:obs}).
        Out of the the LOFAR observations made by~\citet{2021A&A...645A..59T}, shown as coloured line segments, the `bursty' L569131 signal is contemporaneous with the magnetic observations while the `slowly varying' L570725 signal was observed about two weeks later.
        The bursty signal was observed near quadrature while the slowly varying signal was observed near the inferior conjunction of \(\tau\)~Boötis~Ab.
    }\label{fig:phases}
\end{figure}

In this work we present a new magnetic field map of \(\tau\)~Boötis~A that is contemporaneous with the radio observations of~\citet{2021A&A...645A..59T}. The map is constructed based on Zeeman-Doppler imaging \citep[ZDI,][]{1989A&A...225..456S}, see also~\citet{2009ARA&A..47..333D}.
To create the magnetic field map, we conducted spectropolarimetric observations of \(\tau\)~Boötis~Ab  from 15--22 February 2017, which is contemporaneous with the bursty LOFAR signal and nearly contemporaneous with the slowly varying signal, as indicated in Fig.~\ref{fig:phases} and in Table~\ref{tab:obs}.
The phases and orbital configuration are based on the time of the conjunction value in Table~\ref{tab:properties}.

Our spectropolarimetric observations were obtained with the NARVAL instrument~\citep{2003EAS.....9..105A} at the Télescope Bernard Lyot. NARVAL is a spectropolarimeter with a resolution \(R  \sim  65000\) that covers a wavelength range from \SIrange{370}{1050}{\nano\meter}. During the observations, we simultaneously recorded the circularly polarised Stokes~\(V\) spectrum and the total intensity Stokes~\(I\) spectrum.
We obtain a polarimetric sequence from four consecutive sub-exposures, each of which is taken with a different rotation of the retarder waveplate of the polarimeter relative to the optical axis. The Stokes~$I$ spectrum is computed by summing the four sub-exposures, while the Stokes~$V$ spectrum from the ratio of sub-exposures with orthogonal polarisation states. A polarimetric sequence lasts \qty{2400}{\second} and comprises \num{4} sub-exposures of \qty{600}{\second}.

We reduced our data with the \textsc{libre-esprit} pipeline~\citep{1997MNRAS.291..658D}. At \qty{650}{\nano\meter}, the signal-to-noise ratio per instrument pixel of the circularly polarised spectrum ranges from \num{1295} to \num{1559} with an average of \num{1437}.
We used a normalising wavelength of \qty{650}{\nano\meter}, and a normalising Landé factor of \num{1.195}.

We applied the least square deconvolution technique~\citep[LSD,][]{1997MNRAS.291..658D,2010A&A...524A...5K} to combine circularly polarised line profiles into a single LSD profile with a high signal-to-noise ratio for each observation.  We then combined the set of LSD profiles corresponding to different observations by applying the maximum entropy image reconstruction~\citep{1984MNRAS.211..111S} to create the magnetic field map, represented by a set of spherical harmonics coefficients~\citep{1999MNRAS.305L..35J,2006MNRAS.370..629D} up to degree \(\ell_\text{max}=15\). We find an average (maximum) surface magnetic field strength of \qty{1.6}{\gauss} (\qty{5.9}{\gauss}) which is a reduction of \qty{\sim40}{\percent} compared to the observations of~\citet{2018MNRAS.479.5266J} and within the range reported by~\citet{2016MNRAS.459.4325M}. We find that the field is dominantly \qty{74}{\percent} poloidal with \qty{40}{\percent} of the energy in the dipolar field. The level of axisymmetry is \qty{52}{\percent}. These values are comparable to previous observations~\citep[see][]{2018MNRAS.479.5266J}. The vector magnetic map, which shows the radial, azimuthal, and meridional field, is given in appendix~\ref{sec:vector-magnetogram}.

Fig.~\ref{fig:map} shows the radial magnetic field component. The minimum angular size of reproduced features in the magnetic field map is \(\qty{\sim 180}{\degree}/\ell_\text{max} = \qty{12}{\degree}\), but most of the energy is in lower modes, with \qty{90}{\percent} of the energy in modes \(\ell \leq 5\) and \qty{99}{\percent} of the energy in modes \(\ell \leq 8\), giving minimum feature extents of \(\qty{\sim 36}{\degree}\) and \(\qty{\sim 23}{\degree}\) respectively (see Table~\ref{tab:wind-aggregate-results}); this is similar to the Zeeman-Doppler imaging-based magnetic maps we have previously used, e.g., in~\citet{2023MNRAS.524.2042E}.
We see that the northern hemisphere is dominated by positive radial magnetic field values; this agrees with both the \qty{118}{\day} and the \qty{360}{\day} polarity switch cycles considered by~\citet{2018MNRAS.479.5266J}.

\begin{figure}
    \centering
    \includegraphics{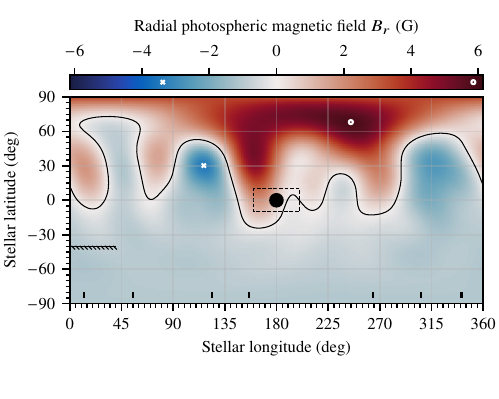}
    \caption{
    Radial surface magnetic  field of \(\tau\)~Boötis~A based on spectropolarimetric observations conducted contemporaneously with the detected LOFAR radio signal.
    The rotation axis of \(\tau\)~Boötis~A is inclined by about \qty{40}{\degree} to the line of sight, hence stellar latitudes below \qty{-40}{\degree} are minimised by the ZDI regularisation.
    The radial field strength ranges from \qtyrange{-3.38}{5.85}{\gauss} and the mean value \(\langle|B_r|\rangle = \qty{1.21}{\gauss}\). The location of the minimum and maximum values are indicated by a white cross and a white circle.
    The subplanetary point on the stellar surface is indicated by a black dot. The dashed square around the subplanetary position indicates the longitude and latitude range that will be used for computing quantities at the position of the planet.
    The phase of the spectropolarimetric observations are indicated by the short tick marks near the longitude axis.
    Note that the full vector magnetogram is shown in Fig.~\ref{fig:vector-magnetogram}.
    }\label{fig:map}
\end{figure}

\section{Stellar wind models}\label{sec:stellar-wind-models}
In this section we briefly describe the wind modelling methodology and the parameter choices we make in this work in order to create several wind models contemporaneous with the radio observations. We describe the resulting wind models and their relative differences.

\subsection{Wind model software}
In this work we obtain steady-state wind solutions by numerically solving the two-temperature magnetohydrodynamic (MHD) equations, extended with heating and cooling terms and two further equations describing the propagation of Alfvén waves energy~\citep[AWSoM model,][]{2013ApJ...764...23S,2014ApJ...782...81V}. Our modelling approach differs from the one used by \citet{2012MNRAS.423.3285V,2016MNRAS.459.1907N}, who simulated the  wind of \(\tau\)~Boötis~A using a polytropic wind model.

Our model is driven by the surface magnetic map shown in Fig.~\ref{fig:map}. As is common in wind modelling, this work only makes use of the radial component of the magnetic field shown in Fig~\ref{fig:map}; the polar and azimuthal components of \(\vec B\) are not fixed, but balanced with the magnetohydrodynamic forces inside the solution domain.

The AWSoM model has been validated for the Sun~\citep{2015MNRAS.454.3697M,2019ApJ...872L..18V,2019ApJ...887...83S}. Many studies have used the model to study the winds of other stars~\citep[e.g.][]{2016A&A...588A..28A,2016A&A...594A..95A,2017ApJ...835..220C,2017ApJ...843L..33G,2021MNRAS.504.1511K,2021MNRAS.506.2309E,2022MNRAS.510.5226E,2023MNRAS.524.2042E,2022MNRAS.509.5117S,2023MNRAS.522..792M,2024MNRAS.529L.140E,2024MNRAS.533.1156S}.
The \awsom{} model is a component in the block-adaptive tree solarwind Roe upwind scheme~\citep[\batsrus{}, ][]{1999JCoPh.154..284P,2012JCoPh.231..870T} code, itself a part of the space weather modelling framework~\citep[\swmf{}, ][]{2005JGRA..11012226T,2012JCoPh.231..870T,2021JSWSC..11...42G}. The model equations and parameters used in this work are described in detail in~\citet{2021MNRAS.506.2309E,2022MNRAS.510.5226E}.

The \awsom{} module  numerically computes the wind density \(\rho\), the wind speed in the stellar frame \(\vec u\), the magnetic field \(\vec B\), the wind thermal pressure \(p_\Thermal\), as well as Alfvén wave energy fluxes propagating along the magnetic field lines. The model also includes the effects of electron heat conduction and radiative cooling.
The model extends from the stellar chromosphere to several times the orbital distance of \(\tau\)~Boötis~Ab. The model assumes that the corona is heated by the dissipation of Alfvén wave energy, which is prescribed at the inner boundary of the model, where the Alfvén wave energy input \(\Pi_\Alfven\) is proportional to the local value of the magnetic field strength \(B_r\) shown in Fig.~\ref{fig:map}. This is modelled in terms of a proportionality constant
\(\Pi_\Alfven/B\). For solar models the value \(\Pi_\Alfven/B = \qty{1.1e6}{\watt\per\square\meter\per\tesla}\) is taken as the default~\citep{2018LRSP...15....4G}. In this work we consider the solar value and a value ten times larger as can be seen in Table~\ref{tab:models} and further expanded below.
The wind models presented here were created with the open source version of the SWMF\footnote{The open version of the space weather modelling framework can now be found on-line at
\url{https://github.com/SWMFsoftware/SWMF}.
}.

\subsection{Model parameters}
As in~\citet{2018LRSP...15....4G} and ~\citet{2021MNRAS.506.2309E,2022MNRAS.510.5226E,2023MNRAS.524.2042E} we use solar chromospheric temperature \(T=\SI{5e4}{\kelvin}\) and number density values \(n=\SI{2e17}{\per\cubic\meter}\) such that the mass density is \(\rho = \qty{3e-10}{\kilogram\per\cubic\meter}\) at the inner boundary of the model. The radius, mass, and rotation period of \( \tau\)~Boötis~A are taken from Table~\ref{tab:properties} and the references therein.

\begin{table}
    \caption{Model cases considered in this work. Scaling is applied to two parameters; the radial surface magnetic field strength \(B_r\) as shown in Fig.~\ref{fig:map} and the Alfvén flux-to-field ratio \(\Pi_\Alfven/B\) of Section~\ref{sec:stellar-wind-models}. All other modelling parameters are fixed for the four cases.}\label{tab:models}
    \centering
    \begin{tabular}
        {l
        l
        S[table-format=3.3e0]
        S[table-format=1.1e1]
        }
    \toprule
    Model & ID & {\(\langle|B_r|\rangle\)} & {\(\Pi_\Alfven/B\)} \\
          &    &
          {\(\big(\unit{\gauss}\big)\)}
          &
          {\(\big(\unit{\watt\per\square\meter\per\tesla}\big)\)} \\ \midrule
    Unscaled model                        & B1   & 1.21 & 1.1e6 \\
    \(10\times\) scaled magnetic field    & B10  & 12.1 & 1.1e6 \\
    \(100\times\)  scaled magnetic field  & B100 & 121  & 1.1e6 \\
    \(10\times\) scaled Alfvén wave flux  & SA10 & 1.21 & 1.1e7 \\
    \bottomrule
    \end{tabular}
    \end{table}

To account for uncertainties in the choice of model parameters, and to study the effect of scaling key wind modelling parameters, we consider four different wind models.
As the magnetic field strength found with ZDI is often an underestimate~\citep{2015ApJ...813L..31Y,2018MNRAS.480..477V,2019MNRAS.483.5246L,2020A&A...635A.142K,2020ApJ...894...69S}, see also the discussion in \citet{2021MNRAS.506.2309E}, we consider the ZDI reported field strength to be a lower bound, and create three wind models corresponding to scaling the magnetic field by a factor of 1, 10, and 100. In the following we refer to these models at as the B1, B10, and B100 models. We note that while a \(10\times\) scaling of the magnetic field may be appropriate~\citep[e.g.\  to match spin-down torques as in][]{2024MNRAS.529L.140E}, the \(100\times\) scaling of the B100 model is created to study the scaling of radio flux densities with increasing \(B\). Taken on its own, the B100 model should not be considered representative of the steady-state wind. We note, however, that \citet{1999JGR...10414025F} suggested that transient wind variations could increase the radio flux intensity by a factor of \numrange{e2}{e3}. The B100 model could be representative of the wind conditions giving rise to such as surge in radio flux intensity.

An increased Poynting flux results in a denser stellar wind~\citep{2020A&A...635A.178B}, and this parameter has been scaled when modelling young stars~\citep{2021ApJ...916...96A}. To better understand how the stellar surface Poynting flux would influence radio emissions, we also create the SA10 model, in which the Poynting flux-to-field value is increased by a factor of 10 (and \(B\) is unscaled). The parameters that we vary between our models are given in Table~\ref{tab:models}.

\subsection{Global wind model results}

Table~\ref{tab:wind-aggregate-results} gives aggregate quantities computed from the magnetic map and the resulting wind models. These are provided to enable comparison with other modelling work, in particular~\citet{2023MNRAS.524.2042E}.
\begin{table*}
    \caption{
        Aggregate parameters computed from surface magnetic map and the resulting wind models. The `Solar' column refers to the Sun-G2157 solar maximum model of \citet{2023MNRAS.524.2042E} and is included for comparison (note that thermal bremsstrahlung was not computed for the Sun-G2157 model). For the wind mass- and angular momentum loss rates we have used the values \(\dot M_\Sun\approx\qty{1e9}{\kilogram\per\second}\)~\citep[e.g.][]{2021LRSP...18....3V} and \(\dot J_\Sun \approx \qty{2e23}{\newton\meter}\)~\citep[e.g.][]{2019ApJ...883...67F}.
    }\label{tab:wind-aggregate-results}

    \begin{tabular}{
        lc
        S[table-format=3.3]
        S[table-format=3.3]
        S[table-format=3.3]
        S[table-format=3.3]
        S[table-format=3.3]
        }
        \toprule
        Parameter                                   &                                                                & {B1}                   & {B10}                   & {B100}                  & {SA10}                 & Solar \\
        \midrule
        Mean surface radial field strength       & {\(\big(\unit{\gauss}\big)\)}            &            1.19 &            11.9 &             119 &            1.19 &            5.55  \\
Max surface radial field strength        & {\(\big(\unit{\gauss}\big)\)}            &            5.74 &            57.4 &             574 &            5.74 &            91.3  \\
Unsigned radial flux at surface          & {\(\big(\qty{e15}{\weber}=\qty{e23}{Mx}\big)\)} &            1.54 &            15.4 &             154 &            1.54 &            3.38  \\
Mean surface field strength              & {\(\big(\unit{\gauss}\big)\)}            &            1.69 &            16.9 &             170 &            1.68 &            9.26  \\
Surface dipole     energy fraction       & {\(-\)}                                  &           0.371 &           0.374 &           0.379 &            0.37 &           0.006  \\
Surface quadrupole energy fraction       & {\(-\)}                                  &           0.173 &           0.173 &           0.172 &            0.17 &          0.0329  \\
Surface magnetic energy \qty{90}{\percent} min. feature extent & {\(\big(\unit{\degree}\big)\)}           &              36 &              36 &              36 &              36 &            12.9  \\
Surface magnetic energy \qty{99}{\percent} min. feature extent & {\(\big(\unit{\degree}\big)\)}           &            22.5 &            22.5 &            22.5 &            22.5 &              12  \\
Open magnetic flux fraction              & {\(-\)}                                  &           0.485 &           0.326 &           0.229 &           0.727 &           0.468  \\
Open fieldline surface fraction          & {\(-\)}                                  &           0.336 &            0.22 &           0.132 &           0.411 &             0.05  \\
Magnetic dipole inclination (obliquity)  & {\(\big(\unit{\deg}\big)\)}              &            28.5 &            26.9 &              27 &            29.5 &            66.9  \\
Average Alfvén radius                    & {\(\big(\unit{\Rstar}\big)\)}            &            6.98 &            13.1 &              20 &            3.73 &            6.86  \\
Average Alfvén radius in equatorial plane & {\(\big(\unit{\Rstar}\big)\)}            &            5.35 &            9.88 &            15.6 &            2.87 &            5.36  \\
Wind mass loss rate                      & {\(\big(\qty{e9}{\kilogram\per\second}\approx\dot M_\Sun\big)\)} &             2.1 &            18.7 &             143 &            28.4 &            3.84  \\
Wind angular momentum loss rate          & {\(\big(\qty{e24}{\newton\meter}  \approx 5 \dot J_\Sun \big)\)} &            1.28 &            33.8 &             480 &            5.31 &            2.14  \\
        Thermal bremsstrahlung intensity at \qty{10}{\mega\hertz}          & {\(\big(\qty{e-9}{\jansky}\big)\)} &            0.32 &            1.83 &             10.66 &            0.76 &  {---}            \\
        Thermal bremsstrahlung intensity at \qty{30}{\mega\hertz}          & {\(\big(\qty{e-9}{\jansky}\big)\)} &            1.25 &            6.73 &             52.93 &            3.32 &  {---}            \\
        \bottomrule
    \end{tabular}
\end{table*}
In Table~\ref{tab:wind-aggregate-results}, the mean and maximum unsigned surface radial field, and the unsigned surface radial flux are computed directly from the magnetic map of Fig.~\ref{fig:map}.
The mean surface field strength, surface field dipole and quadrupole energy fractions, and
the surface energy minimum feature extents~\citep[see][]{2022MNRAS.510.5226E} are computed at the stellar surface, accounting for the magnetic field in the steady-state wind models, i.e. the radial field is as in Fig.~\ref{fig:map} and the perpendicular field values that have settled to their steady state values.
The open magnetic flux fraction, open field-line surface fraction, magnetic dipole inclination, average Alfvén radii, and wind mass- and angular momentum-loss rates are computed from the full, steady-state wind model solution as in~\citet{2021MNRAS.506.2309E}.

The inclination of the magnetic dipole axis with respect to the axis of rotation is between \qtyrange{27}{30}{\degree} for the four models. The average Alfvén radius (see Appendix~\ref{sec:alfven-surface}) increases with increasing field strength from models B1 to B100 and is the smallest in the SA10 model. We see that the wind mass loss rate is comparable with solar maximum in the B1 model, and increases with both the magnetic field scaling factor and the Poynting flux scaling factor.
The angular momentum loss rates also increase with increasing \(B\) and increasing Poynting flux-to-field ratio \(\Pi_\Alfven/B\) as expected~\citep[e.g][]{2020A&A...635A.178B,2023MNRAS.524.2042E,2024MNRAS.529L.140E}.

We note that \citet{2010A&A...515A..98P} measured an X-ray flux of \qty{3.21e-12}{\erg\per\square\centi\meter\per\second} for \(\tau\)~Boötis. This is equal to a surface X-ray flux of \qty{7.9e5}{\erg\per\square\centi\meter\per\second}. The relation of~\citet{2021ApJ...915...37W} suggests that \(\tau\)~Boötis~A loses mass via wind at a rate of \(\dot{M}\simeq 20 \, \dot{\mathrm M}_\odot\). This value is roughly comparable to the B10 case where \(\dot{M}=\num{18.7}\,\dot{\mathrm M}_\odot\), (\(\dot{\mathrm{M}}_\Sun\approx \qty{1e9}{\kilogram\per\second}\) is the solar mass-loss rate).

\begin{figure*}
    \centering
    \includegraphics{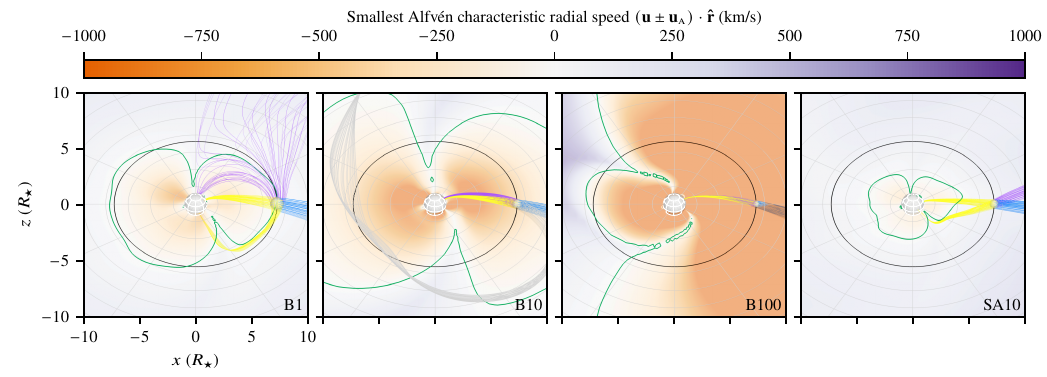}
    \caption{
        Alfvén characteristic flow lines \(\vec u \pm \vec u_\Alfven\) that intersect the planetary magnetosphere, represented by a sphere of radius \(R_\Mag\).
        The purple and blue curves correspond to
        flow lines that {\it originate} at the the planetary magnetosphere, while
        the yellow and grey curves correspond to flow lines that {\it terminate} at the planetary magnetosphere.
        Alfvén wave energy thus flows away from the planet along the blue/purple curves, and towards the planet along the yellow/grey curves.
        A graticule (white) has been added to help visualise the orbital plane and its inclination to the observer.
        The equatorial plane is coloured by the smallest value of the radial component of \(\vec u \pm \vec u_\Alfven\).
       The planetary orbit is indicated as a black ellipse.
       The intersection of the Alfvén surface and the planetary orbital plane is shown as a green curve.
    }\label{fig:wind-z}
\end{figure*}

Fig.~\ref{fig:wind-z} shows key geometric phenomena for the four model cases of Table~\ref{tab:models}.
In this work we apply a coordinate system centred on the star, where the \(xz\)-plane coincides with the plane of sky and the \(y\) axis is directed into the paper plane. The observer is situated in the \(-y\) direction, looking towards \(+y\). The stellar axis of rotation lies in the \(xy\)-plane and is inclined by \qty{40}{\degree} to the line of sight \(\uvec y\), which corresponds to the rotational inclination of \(\tau\)~Boötis A.
We show the system near quadrature, which is the phase of the bursty signal detection. The planet is situated in the \(+x\) direction.

The orbit of \(\tau\)~Boötis~Ab is indicated as a black curve, and the planet's estimated magnetospheric size is shown (to size) at the position of quadrature.  The intersection between the Alfvén surface (see Appendix~\ref{sec:alfven-surface}) and the orbital plane of \(\tau\)~Boötis~Ab is shown as a green curve. We see that \(\tau\)~Boötis~Ab is subalfvénic (inside the Alfvén surface) in the B10 and B100 cases, superalfvénic (outside the Alfvén surface) in the SA10 case, and very close to the Alfvén surface (sometimes called transalfvénic) in the B1 case. The ratio between the wind speed and the Alfvén speed has implications for the shape of the wind-planet interaction region, as discussed in section~\ref{sec:estimated-magnetosphere}.

In Table~\ref{tab:wind-aggregate-results}, we also report radio flux intensity from thermal bremsstrahlung at \qty{10}{\mega\hertz} and \qty{30}{\mega\hertz} following the approach of \citet{2019MNRAS.483..873O,2021ascl.soft01004O}; we provide details about this in Appendix~\ref{sec:brems-and-gyro}. Here we simply record that the computed bremsstrahlung flux intensities are 7--9 orders of magnitude fainter than the \citet{2021A&A...645A..59T} observations.

\subsection{Quantities at the planetary position}
To model the radio emission from the planet \(\tau\)~Boötis~Ab we need to know the wind parameters at the location of the planet, which requires a model of the planetary orbit. While the planetary orbital phase is known to high accuracy (see Table~\ref{tab:properties}), the rotation period of the star is not as accurately known. Furthermore the differential rotation observed using ZDI~\citep{2016MNRAS.459.4325M} introduces some uncertainty about the magnetic field configuration at the sub-planetary point on the star. To account for this we compute planetary quantities in region of azimuthal range \(\pm 20\) degrees and a polar range \(\pm 10\) degrees around the expected planetary position. The resulting values and their range
are reported in Table~\ref{tab:results}. This region is also indicated by the dashed rectangle in Fig.~\ref{fig:map}.
\begin{table*}
    \caption{
        Stellar wind and derived emission-related quantities from the four wind models presented in this work.
        The listed quantities and their variation were computed in the region indicated by a dashed square in Fig.~\ref{fig:map}.  All quantities are computed in the planetary frame unless otherwise stated.
        Note that the wind model assumes a fully ionised hydrogen plasma, so the H$^+$ number density equals the electron number density.
    }\label{tab:results}
    \centering
    \sisetup{
        table-alignment-mode = format,
        table-number-alignment = center
    }
    \begin{tabular}{
        lcc
        S[table-format=4.4(2)]
        S[table-format=4.4(2)]
        S[table-format=4.4(2)]
        S[table-format=4.4(2)]
        }
        \toprule
        Parameter              &                     & Unit                                                               & {B1}                   & {B10}                   & {B100}                  & {SA10}  \\
        \midrule
        Wind speed                     & \(v         \) & $ \big( \qty{e6}{\meter\per\second}\big) $ &  0.453(25) &  0.445(11) &  0.558(20) &  0.440(33)\\
Wind speed (stellar inertial frame) & \(u         \) & $ \big( \qty{e6}{\meter\per\second}\big) $ &  0.418(28) & 0.4041(85) &  0.521(14) &  0.402(40)\\
Density                        & \(\rho      \) & $ \big( \qty{e-18}{\kilo\gram\per\cubic\meter}\big) $ &   5.70(79) &  53.9(4.5) &                                                                                              248(37) &     77(14)\\
Magnetic field strength        & \(B         \) & $ \big( \qty{e-6}{\tesla}\big) $ &  1.159(41) &   7.62(27) &  65.4(3.8) &  1.988(61)\\
Thermal pressure               & \(p_\Thermal\) & $ \big( \qty{e-6}{\pascal}\big) $ & 0.0560(13) & 0.1386(81) &   1.97(41) &  0.477(72)\\
Magnetic pressure              & \(p_\Magnetic\) & $ \big( \qty{e-6}{\pascal}\big) $ &  0.535(37) &  23.1(1.6) &                                                                                            1707(196) &  1.574(95)\\
Ram pressure                   & \(p_\Ram    \) & $ \big( \qty{e-6}{\pascal}\big) $ &  1.157(35) &  10.67(64) &     77(13) &  14.6(1.0)\\
Alfvén pressure                & \(p_\Alfven \) & $ \big( \qty{e-6}{\pascal}\big) $ & 0.2291(91) &  0.722(40) &   2.87(51) &   3.10(14)\\
Magnetopause radius            & \(R_\Mag    \) & $ \big( \unit{R_\Planet} \big) $ &  8.067(27) &  4.736(40) &  2.389(44) &  5.706(46)\\
Obstacle projected area        & \(A_\text{obs}\) & $ \big( \qty{e18}{\meter\squared}\big) $ & 1.1227(76) & 0.3870(65) & 0.0985(36) & 0.5618(90)\\
\midrule
Sound speed                    & \(v_\textsc{s}\) & $ \big( \qty{e6}{\meter\per\second}\big) $ & 0.1288(81) & 0.0655(20) & 0.1147(34) & 0.1021(49)\\
Alfvén speed                   & \(v_\Alfven \) & $ \big( \qty{e6}{\meter\per\second}\big) $ &  0.437(41) &  0.928(71) &   3.75(49) &  0.205(20)\\
Fast speed                     & \(v_\text{fast}\) & $ \big( \qty{e6}{\meter\per\second}\big) $ &  0.455(42) &  0.931(70) &   3.75(49) &  0.229(19)\\
Plasma-\(\beta\)               & \(\beta     \) & $ \big( \unit{1} \big) $ & 0.1051(71) & 0.00604(72) & 0.00120(38) &  0.306(63)\\
Alfvénic Mach number           & \(M_\Alfven \) & $ \big( \unit{1} \big) $ &  1.043(51) &  0.482(29) &  0.151(20) &   2.16(13)\\
Fast Mach number               & \(M_\text{fast}\) & $ \big( \unit{1} \big) $ &  0.999(47) &  0.480(29) &  0.151(20) &  1.925(95)\\
\midrule
Electron density               & \(n_\text{e}\) & $ \big( \qty{e9}{\per\cubic\meter}\big) $ &   3.41(47) &  32.3(2.7) &                                                                                              148(22) &  45.9(8.4)\\
Plasma frequency               & \(f_\Plasma \) & $ \big( \qty{e6}{\hertz}\big) $ & 0.0767(52) & 0.2363(99) &  0.506(38) &  0.281(24)\\
Electron gyrofrequency         & \(f_\Gyro   \) & $ \big( \qty{e6}{\hertz}\big) $ & 0.0324(11) & 0.2132(76) &   1.83(11) & 0.0557(17)\\
\midrule
Polar cap colatitude           & \(          \) & $ \big( \unit{\degree} \big) $ &  7.121(24) &  12.19(10) &  24.76(49) & 10.094(83)\\
\midrule
Perpendicular magnetic field strength & \(B_\perp   \) & $ \big( \qty{e-6}{\tesla}\big) $ &  0.798(28) &   4.89(12) &  28.8(1.6) &  1.441(97)\\
Wind flow-magnetic field angle & \(\angle (\vec{v},\vec{B})\) & $ \big( \unit{\degree} \big) $ &  43.6(2.2) &  40.0(1.1) &  26.2(2.1) &  46.7(4.5)\\
\bottomrule

    \end{tabular}
\end{table*}

Where applicable, the quantities in Table~\ref{tab:results} are computed in the planetary frame. For example, the quantity \(\vec v\) is the wind speed in the planet's frame, so that \(\vec v = \vec u - \vec u_\Planet\) where \(\vec u\) is the wind velocity and \(\vec u_\Planet\) is the planet's orbital velocity in stellar inertial frame.
This gives a Keplerian velocity \(u_\Planet = \sqrt{GM_\Star/a_\Planet}=\qty{161}{\kilo\meter\per\second}\), where \(a_\Planet\) is the semi-major axis of the planet's orbit (see Table~\ref{tab:properties}).
The direction of \(\vec u_\Planet\) is such that \(\uvec r_\Planet \times \uvec u_\Planet = \uvec \Omega\), where \(\vec r_\Planet\) is the vector pointing from the star to the planet. Hence, at every point the planet is assumed to be travelling directly in the direction of increasing longitude. In our analysis, we assume that the orbit of \(\tau\)~Boötis~Ab is such that the sub-planetary point on the star is stationary. (In this work the caret or `hat' symbol denotes the unit vector, i.e., \(\uvec a\) is the unit vector in the \(\vec a\) direction.)

In the following we discuss the computation of key quantities used to model the radio emission from \(\tau\)~Boötis~Ab. These key quantities are also given in Table~\ref{tab:results}, along with some intermediate values used in their computation. These intermediate quantities have their usual meaning.

\subsection{Estimated magnetospheric properties}\label{sec:estimated-magnetosphere}
A planetary magnetosphere is the region around a planet which is dominated by the planetary magnetic field~\citep{1930Natur.126..129C}.
The size of the upstream magnetosphere can be estimated by balancing the total (i.e. thermal, kinetic and magnetic) pressure of the stellar wind \(p_\Wind\) with the pressure of the planetary magnetic field \(p_\Planet\)~\citep{1931TeMAE..36...77C, 2009ApJ...703.1734V}.
The total wind pressure is calculated at the location of the planet, and in the frame of the planet, from the steady-state numerical wind model~\citep[as in e.g.][]{2011AN....332.1055V}
\begin{equation}\label{eq:wind-pressure}
    p_\Wind = p_\Thermal + |\vec B|^2/(2\mu_0) + \rho |\vec v|^2,
\end{equation}
where the right hand side terms are the thermal pressure \(p_\Thermal\), magnetic pressure \(p_\Magnetic\), and ram pressure \(p_\Ram\), respectively.
It is sometimes assumed that the wind ram pressure dominates the \(p_\Wind\). This holds in the superalfvénic regime (case B1, SA10) but in the subalfvénic regime (case B10, B100) magnetic pressure is the dominant term in the wind pressure (see Table~\ref{tab:results}).

By assuming that the planet's pressure is magnetically dominated and that its field is dipolar, the planetary pressure acting against the stellar wind is
\(p_\Planet \simeq \big(B_\Planet (r/R_\Planet)^{-3}\big)^2/(2\mu_0)\),
the distance to the magnetopause \(R_\Mag\) can be estimated by balancing the wind pressure with the planetary magnetic pressure:
\begin{equation}\label{eq:pressure-balance}
    {R_\Mag}/{R_\Planet} \simeq ({\mu_0})^{-1/6} B_\Planet^{1/3} p_\Wind^{-1/6},
\end{equation}
see Appendix~\ref{sec:magnetosphere-size} for a derivation of this equation. Note that an additional term of order unity is sometimes applied to correct the magnetic field near the magnetopause for the effects of electrical currents in the magnetopause~\citep{2004pssp.book.....C}.

When a planet is immersed in a
supercritical\footnote{In this work we use the term {\it supercritical} to refer to wind flowing faster than the fastest wave speed in the medium.}
wind, magnetohydrodynamic waves are unable to propagate upstream. This causes a bow shock to form on the windwards-facing side of the planetary magnetosphere, where the wind material is rapidly decelerated and deflected. Overall, the planetary magnetosphere takes on a teardrop shape with a blunt `nose' and a narrow, elongated tail~\citep{1995isp..book.....K}. The orientation of the teardrop shape is determined by the wind velocity in the planetary frame \(\vec v = \vec u  - \vec u_\Planet\) where \(\vec u_\Planet\) is the planet's orbital velocity in the stellar frame. The effective size of the magnetosphere as an obstacle to the wind is therefore approximately the projected area of a circle with radius \(R_\Mag\), where \(R_\Mag\) is the distance to the magnetopause.

The fastest magnetohydrodynamical wave is aptly called the `fast magnetosonic wave'; it travels at speeds \(u_\Fast\) up to \((u_\Alfven^2 + c_\text{s}^2)^{1/2}\) where
\(\vec u_\Alfven=\vec B/\sqrt{\mu_0 \rho}\) is the Alfvén velocity
and
\(c_\text{s}=\sqrt{\gamma p_\Thermal/\rho}\) is the adiabatic sound speed.
When the magnetic pressure dominates the thermal pressure ( \(p_\Magnetic \gg p_\Thermal\), see eq.~\ref{eq:wind-pressure}), \(u_\Alfven \gg c_\text{s}\), and the fast magnetosonic speed \(u_\Fast\approx u_\Alfven\). This is known as a low-\(\beta\) plasma, where \(\beta = p_\Thermal/p_\Magnetic\) is the ratio of thermal to magnetic pressure. In low-\(\beta\) plasmas it is common to use the term {\it superalfvénic} flow to refer to supercritical flow even though the fast magnetosonic wave speed governs bow shock formation; we follow this convention throughout this work. The values of the parameters discussed here are given in Table~\ref{tab:results}. We see that plasma-\(\beta\) varies from \numrange{0.001}{0.3} in the four models, with the lowest value in the B100 model and the highest in the SA10 model.

In a resolved planetary model, the shape and orientation of the magnetosphere is complex~\citep[see e.g.][]{2015ApJ...815..111S,2024MNRAS.tmp.2263P}. We return to discussing the shape of the magnetosphere when we discuss the resolved planetary magnetohydrodynamic model in  Section~\ref{sec:planetary-mhd}.

\subsection{Alfvén wings}\label{sec:alfven-characteristics}
When the speed of the planet through the wind does not exceed \(u_\Fast\) and thus \(u_\Alfven\), the planet is said to be in the subalfvénic regime. In this scenario magnetohydrodynamic waves can propagate upstream and perturb the stellar wind flow, resulting in gradual deceleration of plasma prior to the magnetopause, whereby the magnetopause itself remains the boundary between stellar and planetary magnetic field. In this case Alfvén waves excited by the interaction form (in the planet's frame of reference) standing waves confined to flux tubes originating from the planetary magnetic poles which are referred to as Alfvén wings~\citep{1965JGR....70.3131D,2016spai.book.....R}. The wings  are regions of enhanced magnetic field strength and plasma density~\citep{1998JGR...10319843N,2013A&A...552A.119S}.

The Alfvén wave energy travels along magnetic field lines with group velocity \(\pm \vec u_\Alfven\),
while the Alfvén waves are simultaneously advected downstream with the wind velocity \(\vec u\). Consequently, the Alfvén wings are inclined with respect to the background (i.e. stellar) magnetic field.
The shape of Alfvén wings can be found from tracing the Alfvén characteristics~\citep[e.g.\ ][]{2018haex.bookE..27S}
\begin{equation*}
    \vec c_\Alfven^\pm = \vec u \pm \vec u_\Alfven = \vec u \pm \vec B/\sqrt{\mu_0 \rho};
\end{equation*}
here \(\vec c_\Alfven^\pm\) are known as the~\citet{1950PhRv...79..183E}~variables. Alfvén waves should be understood as travelling in both the \(\vec c_\Alfven^+\) and \(\vec c_\Alfven^-\) directions in both the subalfvénic and superalfvénic regimes.

Alfvén wings, which connect the planet to the star, are potential channels of star-planet interactions, leading Alfvén wave energy from the planetary magnetosphere towards the stellar surface and causing radio emissions. Fig.~\ref{fig:wind-z} shows the flow lines of Alfvén characteristics that originate at the planetary magnetosphere in blue and purple. A site of potential planet-induced radio emission exists wherever these lines reach the stellar surface. From the figure we see that the transalfvénic B1 case and the superalfvénic B10 and B100 cases each have formed one Alfvén wing connecting the planet to the star. As expected, no Alfvén wing has formed in the superalfvénic SA10 case. In this work we do not model magnetic star-planet interactions of this type further but note that the presence of Alfvén wings is a necessary condition for such interactions to occur.
The yellow and grey curves in Fig.~\ref{fig:wind-z} represent Alfvén characteristics that terminate at the planetary magnetosphere; they are included for completeness.

\section{Radio emissions power}\label{sec:radio-emissions-power}
To estimate the strength of the expected radio signals from \(\tau\)~Boötis~Ab we need to consider the various mechanisms that can produce radio emissions in the system.
Exoplanetary radio emissions are categorised into incoherent mechanisms, such as thermal bremsstrahlung and gyroemission, and coherent mechanisms, such as plasma emissions and electron cyclotron maser emissions, see e.g.~\citet{2004AdSpR..33.2045Z,2008SoPh..253....3N,2024arXiv240915507C}. Space weather models permit the modelling of several of the contributing factors of exoplanetary radio emissions, while other factors are determined by microphysical features of the system and require modelling that goes beyond the magnetohydrodynamical approximation.
Nevertheless, the available energy fluxes to power emissions can be estimated by means of magnetohydrodynamical models.

\subsection{Power budget}
The amount of power that drives the observed radio signals can be estimated based on the distance to the \(\tau\)~Boötis system and the signal bandwidth. Following e.g.~\citet{1999JGR...10414025F}, the emitted power \(P\) is related to the
radio flux density by
\(P = \phi  \Delta f D^2 \Omega\)
where
\(\phi\) is the flux and \(\Delta f\) is the signal frequency range as reported by the radio telescope,
\(D\) is the distance to the source, and \(\Omega\) is the solid angle of emission (required since e.g. ECMI is non-isotropic).
For the \(\tau\)~Boötis system we have \(D=\qty{15.66\pm0.08}{\parsec}\) from Table~\ref{tab:properties}, which gives
\begin{equation}\label{eq:power}
    I=\frac{P}{\Omega}= \phi \Delta f D^2
    \approx
    \left(\frac{\phi}{\qty{e3}{\milli\jansky}}\right)
    \left(\frac{\Delta f}{\qty{1}{\mega\hertz}}\right)
    \,
        \qty{2.3e15}{\watt\per\steradian}
\end{equation}
assuming that the signal intensity is constant with respect to frequency. By inspection, equation~\eqref{eq:power} suggests that the required power per solid angle lies around \(\qty{e16}{\watt\per\steradian}\) for the bursty signal; the observed signal strength, observed frequency range, and the required power per solid angle is given in Table~\ref{tab:power}. In Equation~\eqref{eq:power} we have applied the definition
\(\qty{1}{\jansky} = \qty{e-26}{\watt\per\meter\squared\per\hertz}\).
\begin{table}
    \caption{Observed signal strength \(\phi\), observed frequency range \(\Delta f\), and required signal power per steradian \(I\) for the radio signal detections. The values are computed using eq.~\ref{eq:power} and the distance to the \(\tau\)~Boötis system
    \(D=\qty{15.66\pm0.08}{\parsec}\)
     from Table~\ref{tab:properties}.
     }\label{tab:power}
    \centering
    \begin{tabular}{l
        S[table-format=3(2)]
        S[table-format=1]
        S[table-format=1.2(2)e2]
        }
    \toprule
    Signal          & {\(\phi\)}                            & {\(\Delta f\)}                      & {\(I=P/\Omega\)} \\
                    & {\(\big(\unit{\milli\jansky}\big)\)}  & {\(\big(\unit{\mega\hertz}\big)\)}  & {\(\big(\unit{\watt\per\steradian}\big)\)} \\ \midrule
    L569131         & 890(690:500)                          & 6                                   & 1.25(96:70)e16 \\
    L570725 (peak)  & 430\pm30                              & 9                                   & 9.04\pm0.63e15 \\
    L570725 (mean)  & 190\pm30                              & 9                                   & 3.99\pm0.63e15 \\
    \bottomrule
    \end{tabular}
\end{table}

\subsection{Planetary auroral emissions}
The available energy to power magnetospheric emissions is delivered either by the stellar wind or by internal mass transport in rapidly rotating magnetospheres~\citep[e.g.][]{2021A&A...655A..75S}. Close-in exoplanets are, however, expected to rotate slowly due to possible tidal locking with the host star \citep[see][]{2023A&A...671A.133E}. Therefore we focus on stellar wind-powered, magnetospheric emissions from the auroral regions of \(\tau\)~Boötis~Ab.  The auroral emissions considered in this work are produced by means of the electron cyclotron maser instability~\citep[ECMI,][]{1958AuJPh..11..564T,1975JGR....80.4675S,1979ApJ...230..621W,2006A&ARv..13..229T,2015Natur.523..568H} emissions, which produce anisotropic radio emissions near harmonics of the electron cyclotron frequency, given by
\begin{equation*}
    f_\Gyro = \frac{1}{2\pi}\frac{|q_\Electron|B}{m_\Electron} = \qty{2.8}{\mega\hertz}\left(\frac{B}{\qty{e-4}{\tesla}}\right)
\end{equation*}
where \(q_\Electron\) is the electron charge, \(B\) is the local value of the magnetic field strength, and \(m_\Electron\) is the electron mass.
Radio signals can only propagate through the wind plasma if the plasma frequency
\begin{equation}\label{eq:plasma-frequency}
    f_\Plasma = \frac{1}{2\pi} \sqrt{\frac{q_\Electron^2 n_\Electron}{\epsilon_0 m_\Electron}} = \qty{9.0}{\hertz}\sqrt{\frac{n_\Electron}{\qty{1}{\per\cubic\meter}}}
\end{equation}
is lower than the signal frequency; here \(n_\Electron\) is the electron number density and \(\epsilon_0\) is the electric constant. We note that \(f_\Plasma\lesssim f_\Gyro\) does not hold at the location of the planet in our B1 model (see Table~\ref{tab:results}), suggesting that EMCI emissions may not be able to escape from the location of the planet in the B1 case. See Section~\ref{sec:planetary-mhd} for a discussion on escaping planetary radio emission based on our resolved planetary magnetohydrodynamic model.

Planetary radio emissions can also be inhibited if the corona is optically thick beyond the planetary orbit; we compute the optical thickness of the corona at \qty{10}{\mega\hertz} and \qty{30}{\mega\hertz} and find that the optically thick region of the corona does not extend to the orbit of \(\tau\)~Boötis~Ab, even when accounting for projection effects, see appendix~\ref{sec:brems-and-gyro}.

The incident stellar wind energy flux is the sum of thermal, kinetic, and magnetic energy flux convected through the effective area of the magnetospheric obstacle. Radio observations from Solar system planets suggest that a nearly constant fraction of stellar wind energy incident on the magnetosphere is converted into radio emission generated by the ECMI mechanism~\citep{2007P&SS...55..598Z,2010ASPC..430..175Z,2018A&A...618A..84Z}. As mentioned in the introduction, we adopt two complementary approaches to estimate the radio signal strengths emanating from the auroral regions of
\(\tau\)~Boötis~Ab. The two methods are described in the following subsections.

\subsection{Method A: Unresolved Bode's law-based model}\label{sec:methodA}
In the unresolved Bode's law\footnote{
    We refer to the `radiometric Bode's law'; this scaling relation should be distinguished from the `magnetic Bode's law'~\citep{1978Natur.272..147R}, which relates the magnetic field to angular momentum, and the original `Bode's law'~\citep{1768azkd.book.....B} which suggested a pattern in the orbital radius of the solar system planets.
}-based approach, we use our stellar wind models to compute the kinetic energy flux and magnetic energy flux (more specifically the Poynting flux) at the position of \(\tau\)~Boötis~Ab.
These two energy sources are considered to power radio emissions in the solar system, via a series of steps and energy conversions~\citep{2007P&SS...55..598Z,2010ASPC..430..175Z}. The two resulting `mechanisms' are often referred to as the `radio-kinetic'  and `radio-magnetic' scaling laws in the exoplanetary community.

The radio-kinetic emission mechanism is based on assuming the kinetic energy flux across the planetary magnetosphere is the source of power for radio emissions. In the Bode's law-based model this reduces to the flux
 \(\vec F_\Kinetic = e_\Kinetic \vec v\)
of wind material with kinetic energy density \(e_\Kinetic=\frac12 \rho v^2\) across the windwards facing-magnetosphere. The resulting power is~\citep{2007P&SS...55..598Z}
\begin{equation}\label{eq:kinetic-power}
   P_\Kinetic = \int_{A_\Mag} \vec F_\Kinetic \cdot \uvec n \, \mathrm{d} A \approx \frac{1}{2} \rho v^3 \pi R_\Mag^2,
\end{equation}
where the area of integration \({A_\Mag}\) is the surface of the nose of the magnetosphere and \(\uvec n\) is the outwards pointing normal vector to the surface. The rightmost expression in eq.~\eqref{eq:kinetic-power} is exact when the flux \(\vec F_\Kinetic\) is constant across the magnetosphere,
\(A_\Mag = \pi R_\Mag^2\), where $R_\Mag$ is the radius of the upstream magnetopause, and the magnetosphere nose is pointed directly windwards. The computed values of \(P_\Kinetic\) are given in Table~\ref{tab:planetary_radio}.

In the radio-magnetic emission mechanism the power is assumed to be generated by the interaction between the stellar wind and the planetary magnetic field.
By direct analogy with the radio-kinetic mechanism we would consider a magnetic energy density \(e_\Magnetic=B^2/2\mu_0\) being advected at velocity \(v\), resulting in a magnetic energy flux \(\vec F_\Magnetic=e_\Magnetic \vec v\).
The  available power is, however, computed more accurately based on the Poynting flux
\(
\vec S = \mu_0^{-1} (\vec E \times \vec B)
\)
of electromagnetic energy~\citep[e.g.][]{2007P&SS...55..598Z}.
We apply Ohm's law $\vec E = \vec B \times \vec v$ to write
\begin{equation*}
   \vec S = \mu_0^{-1} (\vec B \times \vec v) \times \vec B =
   \mu_0^{-1} (B^2 \vec v - (\vec B \cdot \vec v)\vec B) =
    \mu_0^{-1}  B^2\vec v_\perp
\end{equation*}
where \(\vec v_\perp = \vec v - (\uvec B\cdot \vec v) \uvec B  \) is the component of the wind velocity perpendicular to the magnetic field.
By integrating \(\vec S\) over the magnetosphere the available power is
\begin{equation*}%
P_\Magnetic = \int_{A_\Mag} \vec S \cdot \uvec n \, \mathrm{d} A \approx \mu_0^{-1} \pi R_\Mag^2 B^2 v_\perp.
\end{equation*}
This result differs from the \(\vec F_\Magnetic\)-based estimate by a factor of \(2\sin\theta\), where \(\theta\) is the angle between the magnetic field and the wind velocity. In our models this difference is between \(2\sin \qty{47}{\degree}=1.5\) for the SA10 model, and \(2\sin \qty{26}{\degree}=0.88\) for the B100 model. The values of \(P_\Magnetic\) are given in Table~\ref{tab:planetary_radio}. Unlike the radio-kinetic mechanism, the radio-magnetic mechanism has been argued to better account for both the emissions of the outer planets and Jupiter's moons in the solar system~\citep[see][]{2018A&A...618A..84Z,2018haex.bookE..22Z}.

For completeness we also compute the thermal power incident on the magnetosphere from the enthalpy \(p/(\gamma-1)\) and pressure \(p\):
\begin{equation*}
P_\Thermal = \int_{A_\Mag}
\left(
    \frac{p}{\gamma-1} + p
\right) \vec v \cdot \uvec n \,\mathrm{d}A \approx
\frac{\gamma p}{\gamma-1}v\,\pi R_\Mag^2,
\end{equation*}
here \(\gamma=5/3\) is the adiabatic index. These values are given in Table~\ref{tab:planetary_radio}; they can be seen to be significantly smaller than \(P_\Kinetic\) and \(P_\Magnetic\).

\begin{figure}
    \centering
    \includegraphics{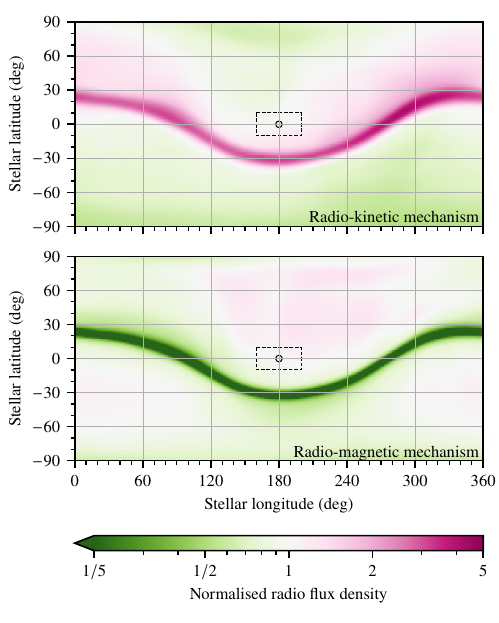}
    \caption{
         Radio flux density variations as a function of planetary position. For the radio-kinetic (top) and radio-magnetic (bottom) mechanisms, the variations shown are computed on a spherical shell of radius equal to the orbital distance of \(\tau\)~Boötis~Ab. In each panel, the values have been normalised to the computed radio flux density at the estimated planetary position. The stellar longitude and latitude values indicate the position of the substellar point.
        Wind speeds are calculated in the planetary frame. In every point the planet is assumed to be travelling in the direction of increasing longitude (this is used for the calculation of \(v\)). The black circle indicates the position of the planet, and the black dashed square indicates the area over which the averages in Table~\ref{tab:results} are calculated.
    }\label{fig:intensity-sphere}
\end{figure}

Our stellar wind model now permits the easy calculation of \(P_\Kinetic\) and \(P_\Magnetic\) at any location, and we see that the asymmetry of the surface magnetic field of \(\tau\)~Boötis~A gives rise to variations in \(P_\Kinetic\) and \(P_\Magnetic\) depending on the exact position of the planet.
Fig.~\ref{fig:intensity-sphere} illustrates the variation in kinetic and magnetic radio flux density for the B10 model; the situation is similar for the other model cases presented in this work.  The values shown are computed from the B10 model in Table~\ref{tab:models} and the assumptions detailed in the previous sections.
All quantities are taken in the planetary frame; in every point, the planet is assumed to be travelling in the direction of increasing longitude. The black circle indicates the position of the planet, and the dashed square shape indicates the region where the values of Table~\ref{tab:results} are computed.
From Fig.~\ref{fig:intensity-sphere} we can see that the available power per solid angle varies considerably. Thus knowledge of the planetary phase, the magnetic map, and the subplanetary point on the star significantly reduces the uncertainty of the reported radio flux intensities given in section~\ref{sec:radio-flux-densities}.
\newcommand{\epsilonreplacement}{\eta'}
\subsection{Method B: Resolved three-dimensional model of the planetary magnetosphere}\label{sec:planetary-mhd}

In the resolved model approach, we use the space weather conditions at the location of \(\tau\)~Boötis~Ab to initialise and drive a resolved magnetohydrodynamic model of the planetary magnetosphere. This permits us to model how the magnetic energy in the planet's auroral regions is generated by the entire stellar wind-planet interaction, which incorporates kinetic, magnetic and thermal energy fluxes.
\begin{figure*}
    \centering
    \includegraphics{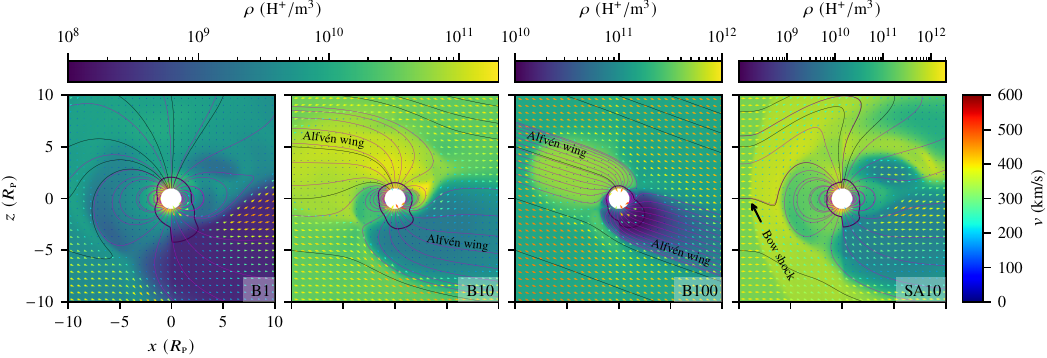}
    \caption{
        Plane $xz$ cross sections of the magnetosphere of \(\tau\)~Boötis Ab in our resolved  three-dimensional model of the planetary magnetosphere (method B). The background colours show the plasma number density \(\text{H}^{+}/\text{m}^3\) according to the panel's top colourbar. Velocity vectors and are indicated by arrows coloured by magnitude according to the right colourbar. Black and magenta lines depict magnetic field line projections into the $xz$-plane. The magnetic dipole moment and rotation vector of the planet are parallel to \(z\). The solid dark violet line indicates the equality between the electron plasma frequency and gyrofrequency, $f_\Plasma/f_\Gyro=1$. Radio emission originating in a region with $f_\Plasma/f_\Gyro<1$, i.e., inside the closed violet curve, may escape the planet's vicinity due to the gyrofrequency exceeding the local plasma frequency. Alfvén wings develop in the subalfvénic cases B10 and B100. A bow shock forms in the superalfvénic scenario SA10. In the transalfvénic scenario B1 neither Alfvén wings nor a bow shock form; the discontinuity visible at $-7.5\,R_\Planet$ is the upstream magnetopause.
    }\label{fig:TauBooMagnetosphere}
\end{figure*}
The model presented here is based on the model of \citet{2023A&A...671A.133E} and uses the same planetary as well as numerical parameters. The resolved planetary simulations are carried out using the PLUTO\footnote{In this work we have made use of PLUTO version 4.4, which is available at \url{https://plutocode.ph.unito.it/}.
}~\citep{2007ApJS..170..228M}
magnetohydrodynamic code. A dipolar planetary magnetic field is assumed with a magnetic moment anti-parallel to the stellar rational axis \(\uvec \Omega\) and a polar magnetic field strength of 9 G \citep{2021A&A...645A..59T}. We assumed a static, spherically symmetric molecular hydrogen atmosphere that acts upon the plasma in terms of sources and sinks of plasma mass, momentum, and energy. We consider photoionisation of neutral particles that is affected by the planetary shadow, charge exchange between ions and neutrals, and dissociative recombination. For a detailed description of the planetary magnetohydrodynamic model and the parametrisation of physical processes, we refer the reader to \citet{2023A&A...671A.133E}.
The spherical simulation box has a radial extent of $70R_\Planet$, centred around the planet. The upstream hemisphere of the box where $x<0$ is enclosed by the inflow boundary where we set the steady state stellar wind parameters.
The steady-state stellar wind boundary conditions are obtained from the stellar wind conditions at the orbit of \(\tau\)~Boötis Ab for each stellar wind model separately (see Table~\ref{tab:results}). In this section, we carry out the simulations in spherical coordinates and present the results in a Cartesian system with the $x$-axis parallel to the stellar wind velocity in the frame of the planet. The $z$-axis is aligned with \(\uvec \Omega\) and the thus with the normal vector of the orbital plane (the stellar rotation vector, planetary rotation vector, planetary orbital vector, and planetary dipole moment are all assumed parallel to $z$).

\subsubsection{Structure of the interaction}
In Fig.~\ref{fig:TauBooMagnetosphere} we show $xz$-plane cross sections of the simulated planetary magnetosphere. The subalfvénic nature of the stellar wind for models B10 and B100 causes Alfvén wings to develop, of which the northern wing ($+z$) connects back to the star, while the southern wing ($-z$) is directed away from the star. The superalfvénic stellar wind in model SA10 leads to the formation of a bow shock with a magnetosheath of thickness about 1.5 $R_\Planet$. The B1 stellar wind model presents a special case in which the wind is slightly superalfvénic ($M_\Alfven\approx1.04$) and slightly sub-fast magnetosonic ($M_\Fast=0.99$). The wind in this scenario is thus transalfvénic in nature in which degenerate Alfvénic wings develop and the upstream stellar wind plasma is perturbed by fast mode waves.
Due to the low stellar wind dynamic pressure, the magnetosphere is large, with a stand-off distance of $R_\Mag\approx 7\,R_\Planet$ in the basic B1 and superalfvénic model SA10. The enhanced stellar wind kinetic and magnetic pressure in the B10 and B100 models cause the magnetosphere to shrink accordingly to $R_\Mag < 3 R_\Planet$. We note that the magnetospheric radii agree well with the \(R_\Mag\) estimates based on equation~\eqref{eq:pressure-balance}, given in Table~\ref{tab:results}.

The simulated magnetospheres present a semi-open (i.e. perpendicular) magnetosphere because the stellar wind magnetic field is nearly perpendicular to the planetary magnetic axis, which is assumed perpendicular to the orbital plane. In comparison with an antiparallel configuration, this geometry results in a weaker magnetic coupling between wind and planetary magnetic field, leading to a reduced transfer efficiency from stellar wind energy towards free magnetic energy in the vicinity of the planet. The transfer of energy would be more efficient with a larger stellar wind magnetic field component antiparallel to the planetary magnetic moment, resulting in an enhanced magnetic reconnection efficiency \citep{2012bspp.book.....B,2023A&A...671A.133E}.

\subsubsection{Plasma frequency in Alfvén wings}
In Fig.~\ref{fig:wind-z} we saw that the Alfvén surface extends past the orbit of \(\tau\)~Boötis~Ab in the subalfvénic models B10 and B100, and the location of the resulting Alfvén wings.
The near-planet behaviour of these wings can be seen in the corresponding cases in Fig.~\ref{fig:TauBooMagnetosphere};
we see a sharp contrast in the plasma density between the
northern ($+z$) Alfvén wing which connects to the star, and
southern ($-z$) Alfvén wing which is directed away from the star. The northern Alfvén wing is nearly aligned with the stellar wind flow \(\vec v\). Consequently, incident plasma accumulates within the wing, where it is decelerated by enhanced magnetic pressure and by collisions with neutrals in the ionosphere. The southern wing, on the other hand, is not subject to direct plasma injection and thus exhibits a reduced plasma density of about two orders of magnitude lower than the northern wing. This strong asymmetry in plasma density between the southern and northern wings is a consequence of the small angle between the stellar wind velocity \(\vec v\) and the magnetic field \(\vec B\), which puts the planetary wake inside the downstream-directed Alfvén wing.

The density variation gives rise to an increased electron plasma frequency (see eq.~\ref{eq:plasma-frequency}) in the northern wing, and an extended low frequency region in the southern wing.  The density is crucial because it determines the local plasma frequency, which will affect radio wave transmission considerably.
For auroral radio emission to be generated, and not to remain trapped near the source it is required that the ECMI emission frequency (which corresponds to the local electron gyrofrequency $f_\Gyro$) exceeds the local electron plasma frequency $f_\Plasma$~\citep[e.g.][]{2006A&ARv..13..229T}. The violet contour line in Fig.~\ref{fig:TauBooMagnetosphere} denotes the outer boundary of the region\footnote{
Note that the spherical coordinates of the resolved model introduce a singularity at the polar axis. This causes a slight jump in physical quantities, seen as a kink in the contour line along the $z$-axis of Fig.~\ref{fig:TauBooMagnetosphere}. This is reflected in the frequency ratio, but does not influence the results of this work.
}
in which \(f_\Plasma / f_\Gyro < 1\).
Within this region, i.e. between the planet and the purple curve, excited radio emission is likely to be generated and to escape from the planet's vicinity \citep[e.g.][]{2017MNRAS.469.3505W}.
Note that \(f_\Plasma/f_\Gyro \ll 1\) is likely required for escape and efficient generation of ECMI waves~\citep{2017GeoRL..44.4439L,2020GeoRL..4790021L,2023pre9.conf03095C,2023JGRA..12831985L,2024JGRA..12932422C,2025esoar.73608390C}. Accounting for this in our analysis would reduce the size of the emission region but not affect the upper limits presented in this work.
We see that in the southern wing the local
gyrofrequency exceeds the local plasma frequency and
thus we expect most detectable radio emission to originate from the southern polar region of \(\tau\)~Boötis Ab where the gyrofrequency is highest.

\subsubsection{Auroral Poynting flux}\label{sec:auroral-poynting-flux}
Recently,~\citet{2023A&A...671A.133E}
argued that the {\it auroral} Poynting flux is an appropriate estimator of the electromagnetic energy available to power auroral emissions. The auroral Poynting flux\footnote{
    In~\citet{2023A&A...671A.133E} the \(S_\Aurora\) and \(P_\Aurora\) quantities are referred to as
     \(S_{\Aurora,\parallel}\) and \(P_{\Aurora,\parallel}\). The efficiency factor labelled \(\epsilonreplacement_\Aurora\)  in this work is referred to as \(\epsilon_{\Aurora,\parallel}\).
} is given by
$
S_\Aurora=\big|\vec{S} \cdot \uvec B_0 \big|
$,
i.e. the magnetospheric Poynting flux projected onto the unperturbed dipolar field $\vec{B}_0$ (the dipolar field is described in Appendix~\ref{sec:dipole}).
The auroral Poynting flux mostly includes Poynting fluxes parallel to the auroral magnetic field lines produced by the stellar wind-planet interaction.

To estimate the available power \(P_\Aurora\), we integrate $S_\Aurora$ over a spherical surface $A_2$ with radius $2R_\Planet$; our choice of radius is based on an analogy with Jupiter where the electron acceleration regions are typically found around that distance~\citep{1998JGR...10320159Z}.
Integrating \(S_\Aurora\) over \(A_2\) gives
\begin{equation*}\label{eq:radio_flux_planetary}
    P_\Aurora
    =
    \int_{A_2} \big|\vec{S}\cdot \uvec B_0\big| \,\mathrm{d}A,
\end{equation*}
By taking the absolute value inside the integral we account for Poynting fluxes travelling both parallel and anti-parallel to the unperturbed magnetic field \(\vec B_0\).
The projection of \(\vec S\) onto \(\vec B_0\) naturally forms auroral ovals~\citep[see fig.~2 in][]{2023A&A...671A.133E}. Lower frequency radio emissions, outside the range considered in this work, may be generated at higher altitudes up to ten planetary radii~\citep{2025esoar.73608390C}, but the choice of  radius at which the \(P_\Aurora\) integral is taken does not greatly affect the value of \(P_\Aurora\) in our models.
The computed values for \(P_\Aurora\) are given in Table~\ref{tab:planetary_radio}.

\section{Radio flux densities}\label{sec:radio-flux-densities}
\begin{table}
    \caption{
        Power values and the resulting radio flux density values from the B1, B10, B100, and SA10 models. In the upper half we tabulate the
        kinetic wind power \(P_\Kinetic\),
        magnetic wind power \(P_\Magnetic\),
        thermal wind power \(P_\Thermal\) (for completeness),
        and auroral power \(P_\Aurora\).
        The middle section shows the corresponding intensity values \(I\) for each mechanism, which are computed using the solid angle of \(\Omega=\qty{1.6}{\steradian}\) and the efficiency factors \(\eta\).
        The lower half of the table shows the resulting radio flux densities \(\phi\) predicted for each mechanism.  The flux densities are computed using a bandwidth of \(\Delta f=\qty{6}{\mega\hertz}\), which corresponds to the observed bandwidth of the bursty signal. The radio flux density values are also given in Fig.~\ref{fig:boxplot-power}.
     }\label{tab:planetary_radio}
    \centering
    \sisetup{per-mode=symbol}
\begin{tabular}{l
    S[table-format=1.4(1)]
    S[table-format=2.3(1)]
    S[table-format=3.2(1)]
    S[table-format=2.2(1)]
}
\toprule
Parameter & {B1} & {B10} & {B100} & {SA10} \\
\midrule
$P_\Kinetic     $ $ \big( \qty{e18   }{\watt               } \big) $ &         0.29(1) &         0.92(6) &          2.1(5) &          1.8(2) \\
$P_\Magnetic    $ $ \big( \qty{e18   }{\watt               } \big) $ &         0.37(3) &          5.1(3) &           82(7) &         0.56(4) \\
$P_\Thermal     $ $ \big( \qty{e18   }{\watt               } \big) $ &        0.071(3) &        0.060(4) &         0.27(7) &         0.29(3) \\
$P_\Aurora      $ $ \big( \qty{e18   }{\watt               } \big) $ &           0.154 &           0.197 &           0.430 &           0.138 \\
\midrule
$I_\Kinetic     $ $ \big( \qty{e15   }{\watt\per\steradian } \big) $ &      0.00184(8) &       0.0057(4) &        0.013(3) &        0.011(1) \\
$I_\Magnetic    $ $ \big( \qty{e15   }{\watt\per\steradian } \big) $ &         0.70(6) &          9.6(6) &         154(12) &         1.05(8) \\
$I_\Aurora      $ $ \big( \qty{e15   }{\watt\per\steradian } \big) $ &           0.010 &           0.012 &           0.027 &           0.009 \\
\midrule
$\phi_\Kinetic  $ $ \big( \qty{      }{\milli\jansky       } \big) $ &        0.131(5) &         0.41(3) &          1.0(2) &         0.80(8) \\
$\phi_\Magnetic $ $ \big( \qty{      }{\milli\jansky       } \big) $ &           50(4) &         684(42) &      10989(883) &           75(5) \\
$\phi_\Aurora   $ $ \big( \qty{      }{\milli\jansky       } \big) $ &           0.685 &           0.880 &           1.920 &           0.614 \\
\bottomrule

\end{tabular}
\end{table}

Several physical steps are required to convert the power of the stellar wind incident on the magnetosphere to  radio flux densities. The conversion includes accounting for the efficiency at which wind power is converted to radio signal power, as well as the radio signal solid angle of emission. Throughout this work, we apply a solid angle of emission of~\qty{1.6}{\steradian} following the example of~\citet{2004JGRA..109.9S15Z} for the Jovian decametric radio emission (see Section~\ref{sec:solid-angle-and-bandwidth}).

Having estimated the available power \(P_\Kinetic\), \(P_\Magnetic\) and \(P_\Aurora\), we account for the efficiency factors \(\eta\) and solid angle of emission \(\Omega\) to find the intensity values and compare them with the source intensity \(I\) values of Table~\ref{tab:power}, i.e.
\begin{equation}\label{eq:three_intensities}
    I_\Kinetic = \eta_\Kinetic P_\Kinetic / \Omega,\quad
    I_\Magnetic = \eta_\Magnetic P_\Magnetic / \Omega, \quad\text{and}\quad
    I_\Aurora=\epsilonreplacement_\Aurora P_\Aurora/\Omega.
\end{equation}

\label{sect:efficiency_factors}We now describe the efficiency factors \(\eta\) used in this work.
For the unresolved, Bode's law-based radio-kinetic mechanism, \citet{2007P&SS...55..598Z} estimated that only a constant fraction \(\eta_\text{k} = \num{e-5}\) of the kinetic power is converted into radio emissions. Similarly, for the unresolved Bode's law-based radio-magnetic mechanism we apply a constant fraction \(\eta_\Magnetic = \num{3e-3}\)~\citep{2018A&A...618A..84Z}, which we consider a best-case value (see Section~\ref{sec:efficiency-factors}).
For the resolved planetary model we use $\epsilonreplacement_\Aurora=\num{e-4}$, which is the efficiency \citet{2023A&A...671A.133E} needed to make the simulated energetics consistent with the observation of \citet{2021A&A...645A..59T}\footnote{
    The meaning of the prime symbol in \(\epsilonreplacement_\Aurora\) is discussed in Section~\ref{sec.scaling-and-transfer}.
}. Observational evidence from Jovian radio emission suggests $\epsilonreplacement_\Aurora$ ranging from \numrange{e-5}{e-4}, see~\citet{2021A&A...655A..75S};
the efficiency and resulting flux densities should thus be seen as optimistic upper bounds.
When \(I \gtrsim I_\Kinetic\), \(I \gtrsim I_\Magnetic\), or \(I \gtrsim I_\Aurora\), we can consider the relevant mechanism able to power the observed radio emissions. For each mechanism the \(I\) value scales linearly with the \(P\) value; the intensity values are given in the middle section of Table~\ref{tab:planetary_radio} for completeness.

\begin{figure}
    \centering
    \includegraphics{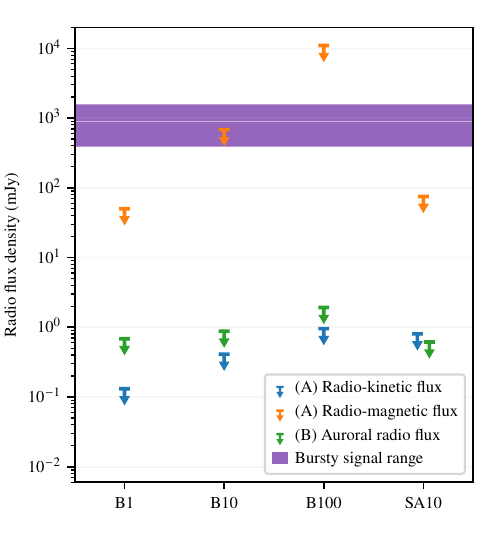}
    \caption{
        Estimated maximum radio flux density at Earth, as observed by LOFAR. The purple band show the flux intensity range reported by LOFAR for the bursty signal. The radio flux densities (downwards-pointing arrows) should be taken as maximum values, as they are based on applying the most favourable efficiency factors \(\eta\) and frequency range \(\Delta f\). The (A) and (B) symbols in the legend refer to the unresolved Bode's law-based approach and the resolved planetary model, respectively. The values shown in this plot are also given in the lower part of Table~\ref{tab:planetary_radio}.
    }\label{fig:boxplot-power}
\end{figure}

We now compute the radio flux densities \(\phi\) that would be observed by LOFAR. As in equation~\eqref{eq:power}, the radio flux density
\(
\phi=I \big/ \big(\Delta f D^2 \big)
\)
where
$\Delta f$ is the radio emission bandwidth of the detections (see Table~\ref{tab:power}), and
\(D\) is the distance to the \(\tau\)~Boötis system (see Table~\ref{tab:properties}).
In the flux density computation, we assume the emission bandwidth to be equal to the bursty signal frequency range \(\Delta f\approx \qty{6}{\mega\hertz}\). The flux densities are thus proportional to the \(I\) values. The results are given in the lower section of Table~\ref{tab:planetary_radio} and plotted in Fig.~\ref{fig:boxplot-power}.

In our models, we find that the radio-kinetic flux density \(\phi_\Kinetic\) is insufficient to power the LOFAR detections in all cases; the highest value, found in the B100 model, is over two orders of magnitude short.
The radio-magnetic flux density \(\phi_\Magnetic\) yields  higher flux density values; it matches the bursty signal range in the B10 case, and exceeds it in the B100 case; the other model cases (B1 and SA10) yield insufficiently powerful radio-magnetic flux densities.
For the auroral radio flux densities \(\phi_\Aurora\) (found with the resolved planetary method), we see that no model case produces a radio flux density inside the range of the bursty LOFAR detection.

\section{Discussion}\label{sec:discussion}
In this work, we have consistently applied best-case efficiency factors, solid angles of emission, and signal frequency ranges in order to obtain an upper bound on the radio flux intensity of planetary emissions from \(\tau\)~Boötis~Ab. Under these best-case conditions, we find that the model B10, in which we apply a \(10\times\) scaling to the stellar surface magnetic field, is able to reproduce the bursty signal with \(\phi=\qty{890(690:500)}{\milli\jansky}\) and bandwidth \(\Delta f=\qty{6}{\mega\hertz}\) by the radio-magnetic Bode's law, and unable to to so in the resolved planetary model (Fig.~\ref{fig:boxplot-power}) or by the radio-kinetic Bode's law.

The radio-kinetic Bode's law-based approach produces \(\phi_\Kinetic\) values that are insufficient to power the LOFAR detections in all the model cases that we consider. One possibility for this could be that the radiomagnetic scaling law is more appropriate to use than the radiokinetic scaling law, as recently suggested by~\citet{2018A&A...618A..84Z}.
We also see (in Fig.~\ref{fig:boxplot-power}) that the unscaled magnetic field model B1, and the SA10 model, in which the stellar surface Alfvén wave flux is scaled by \(10\times\), are not able to reproduce the flux intensity of the bursty signal detection for any of the mechanisms considered in this work. Interestingly, the Alfvén wave flux scaling, which is known to affect e.g. the wind mass loss rate~\citep{2020A&A...635A.178B,2021ApJ...916...96A},
has only a minor effect on the computed radio flux densities.

In our resolved planetary magnetohydrodynamic model (Section~\ref{sec:planetary-mhd}) we find the auroral radio flux ranging from
\qtyrange{0.68}{1.9}{\milli\jansky}.
These radio flux densities are \qty{\sim 250} times lower than those derived in \citet{2023A&A...671A.133E}. In that work, a superalfvénic wind was assumed, based on polytropic, ZDI-driven wind models with ZDI maps of different epochs than the ones presented here  \citep{2016MNRAS.459.1907N}. In Table~\ref{tab:planetary_radio} the Bode's law-based estimates of radio flux density can be seen to scale with magnetic field strength, while the auroral radio flux exhibits a sublinear, non-constant scaling (see Fig.~\ref{fig:boxplot-power}). Interestingly, the signal strengths predicted by the resolved planetary model are comparable in magnitude to the radiokinetic values (see Fig.~\ref{fig:boxplot-power}).

\subsection{The effect of magnetic field scaling on predicted radio fluxes}
With the B1 model, B10 model, and the extreme B100 model, we can investigate the scaling behaviour of the computed radio flux densities with increasing stellar surface magnetic field strength. In Fig~\ref{fig:boxplot-power}, we see that the radio-kinetic flux \(\phi_\Kinetic\) increases by roughly an order of magnitude as the stellar surface magnetic field strength increases by two orders of magnitude. The radio-magnetic flux  \(\phi_\Magnetic\) is more sensitive to the magnetic field and increases by more than two orders of magnitude between the B1 and the B100 case. This can be traced back to small variation in wind speed \(v\) compared to magnetic field strength \(B\) at the planetary location (see the values in Table~\ref{tab:results}).
We find the least flux density variation for the auroral radio flux \(\phi_\Aurora\). The transalfvénic (B1) and superalfvénic (SA10) stellar wind scenarios result in the weakest radio flux densities, with a slightly stronger radio flux density in the transalfvénic (B1) scenario.
With regards to the B1 case, the auroral radio flux densities in the subalfvénic regime (B10 and B100) are enhanced, and their magnitudes appear correlated with the stellar magnetic field. As expected, the strong magnetic fields of the B100 scenario results in the highest radio flux densities.
The low level of variation in the auroral radio flux densities \(\phi_\Aurora\)
which respond even less to increases in the stellar surface magnetic field strength than \(\phi_\Kinetic\)
(Fig~\ref{fig:boxplot-power})
suggest that there is a comparatively weak correlation with stellar wind magnetic energy and a stronger correlation with wind mechanical energy for \(\phi_\Aurora\) (see Section~\ref{sec.transfer}).

\subsection{Scaling behaviour and transfer of energy fluxes in the resolved planetary model}\label{sec.scaling-and-transfer}
The signal intensities predicted by the resolved planetary model do not scale directly with the amount of wind power incident on the planetary magnetosphere. The resolved planetary model permits wind energy to flow by the planet without being converted into auroral Poynting flux despite crossing the planetary magnetosphere.
The fraction of incident power that is converted can be computed in a post-processing step; \citet{2023A&A...671A.133E} refers to this fraction as the {\it transfer function} \(T_\Aurora\). The value of the transfer function depends on the magnetohydrodynamic properties of the incident stellar wind and the planetary magnetic field configuration. To relate \(I_\Aurora\) directly to the wind power at the magnetosphere (here called \(P=P_\Kinetic + P_\Magnetic + P_\Thermal\)) we may thus write
\(\eta_\Aurora = T_\Aurora \epsilonreplacement_\Aurora\) (viz.~eq.~\ref{eq:three_intensities}) to get \(I_\Aurora=\eta_\Aurora P / \Omega\); which is directly comparable to the expressions for \(I_\Kinetic\) and \(I_\Magnetic\) in eq.~\eqref{eq:three_intensities}. The resulting \(\eta_\Aurora\) values are shown in Fig.~\ref{fig:transfer_function}.

\subsubsection{Transfer of energy fluxes}\label{sec.transfer}

\begin{figure}
    \centering
    \includegraphics{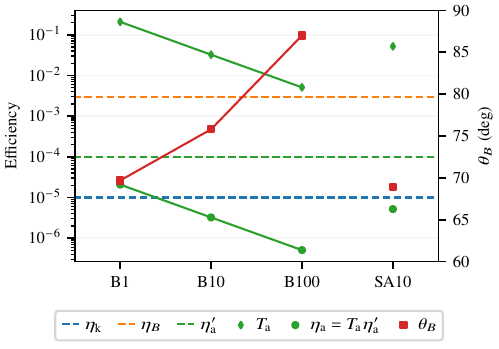}
    \caption{Efficiency factors, transfer function, and angle between planetary magnetic moment and stellar wind magnetic field in the resolved model. The efficiency factors \(\eta_\Kinetic\) and \(\eta_\Magnetic\) (blue/yellow lines) are constant, while the overall auroral radio efficiency \(\eta_\Aurora = T_\Aurora \epsilonreplacement_\Aurora\) (green circles) incorporates the variable transfer function \(T_\Aurora = P_\Aurora / (P_\Kinetic + P_\Magnetic + P_\Thermal)\) (green diamonds) and the constant \(\epsilonreplacement_\Aurora\) (green line).
    \(T_\Aurora\) reflects how the resolved planetary model permits wind energy to flow through the magnetosphere the planet without being converted into auroral Poynting flux; it gives the fraction of converted energy~\citep[see][]{2023A&A...671A.133E}.
    The angle between planetary magnetic moment and stellar wind magnetic field $\theta_\Magnetic$ is shown on a secondary axis as red square symbols. An angle of $\theta_\Magnetic=\qty{0}{\degree}$ signifies an aligned (open) configuration, and $\theta_\Magnetic=\qty{180}{\degree}$ signifies an anti-aligned (closed) configuration.}
    \label{fig:transfer_function}
\end{figure}

For the resolved planetary method in this work, we use a constant auroral radio efficiency $\epsilonreplacement_\Aurora$ (eq. \ref{eq:three_intensities}) motivated from Jovian radio emissions and additionally discuss the transfer function \(T_\Aurora\) (see Fig. \ref{fig:transfer_function}) to further assess the energy transfer towards the magnetosphere and to derive constraints on the auroral radio efficiency.

We calculate the transfer function $T_\Aurora$, which is the fraction of total stellar wind power incident on the magnetosphere transferred to auroral Poynting fluxes (crossing $A_2$, see Sect. \ref{sect:efficiency_factors}). In Fig. \ref{fig:transfer_function} we show the values of the transfer function $T_\Aurora$ for all four stellar wind models together with the angle between stellar wind magnetic field and planetary magnetic moment $\theta_\Magnetic$. The transfer function decreases from approximately \num{0.2} (B1) via \num{0.03} (B10) to \num{0.005} (B100). In the superalfvénic scenario (SA10) $T_\Aurora=\num{0.05}$, which is consistent with the result of \citet{2023A&A...671A.133E}. Our resolved planetary model suggest that the conversion of energy into auroral Poynting fluxes drastically decreases for a subalfvénic interaction, in which the stellar wind energy partition changes from mechanical energy dominated to magnetically dominated. It thus suggests that mechanical energy is the main driver of auroral Poynting fluxes. Even though in the B100 scenario the stellar wind magnetic field has a larger component anti-aligned with the planetary magnetic moment, which typically enhances the reconnection and thus transfer efficiency, this is not enough to compensate the stronger decrease in transfer efficiency.

The derived transfer functions pose a strong upper limit on the overall radio efficiencies $\eta$ (sect. \ref{sect:efficiency_factors} and following) as they dictate how much energy is available within the auroral regions to power electron acceleration and the ECMI mechanism. In the B100 scenario, the overall radio efficiency $\eta_\Aurora=T_\Aurora\epsilonreplacement_\Aurora$ cannot exceed $T_\Aurora=5\times10^{-3}$; indeed \(\eta_\Aurora\) should take on much lower values, since electron acceleration and the ECMI mechanism do not function at efficiencies near unity. This constraint involves both the energy provided by the kinetic and magnetic stellar wind component.

\subsubsection{Scaling behaviour of the estimated radio fluxes}
The scaling behaviour of the auroral radio flux densities \(\phi_\Aurora\) follow the radio-kinetic flux densities \(\phi_\Kinetic\) more closely  than the radio-magnetic flux densities \(\phi_\Magnetic\).
In our resolved planetary models, increasing the magnetic energy of the stellar wind thus plays a comparably insignificant role in enhancing magnetospheric Poynting fluxes, and consequently it has a weak effect on auroral radio emissions. We consider this to be due to the relatively low efficiency of magnetic reconnection because of the weak stellar wind magnetic field component anti-parallel to the planetary magnetic moment \citep[see][]{2012bspp.book.....B}.

In the resolved planetary model, further magnetospheric Poynting fluxes are generated by mechanical perturbations of the magnetosphere. The inductive response of the magnetospheric plasma then generates free magnetic energy that is transported towards the auroral regions via Poynting fluxes. The mechanical energy of the present stellar wind model, however, is orders of magnitude lower compared, for example, to the previous wind model of \(\tau\)~Boötis~Ab \citep{2016MNRAS.459.1907N} used by \citet{2023A&A...671A.133E}.

The geometry of the interaction is poorly constrained and, depending on the orientation of the planetary magnetic field with respect to the stellar magnetic field, the resulting radio flux densities might be enhanced. We previously found that the radio flux densities increase slightly from a semi-open towards an open magnetosphere scenario \citep{2023A&A...671A.133E}; the radio flux densities presented here could be enhanced by approximately a factor of two in a open-magnetosphere (i.e. best case) scenario, in which planetary and stellar magnetic field are perfectly anti-parallel.

In the scope of the radiometric Bode's law, the scaling of radio power with incident Poynting flux is considered the more robust relation \citep{2018A&A...618A..84Z}. In Fig.~\ref{fig:boxplot-power}, however, we observe the radio flux density from our resolved planetary magnetohydrodynamic model to more closely follow the kinetic flux, while the radio flux density obtained from the incident Poynting flux is enhanced by up to two orders of magnitude in the subalfvénic cases (B10 and B100).

The resolved model values are, however, obtained from auroral Poynting fluxes that constitute the maximum available electromagnetic energy that is transported parallel to auroral magnetic field and thus our results are not consistent with the common assumption of the radiometric Bode's law being determined purely by magnetic energy flux, at least in subalfvénic scenarios.

A possible reason to this discrepancy might be the observational bias underlying the radiometric scaling relation whose data is obtained solely from Solar system observations where all planets are exposed to a superalfvénic, but not very energetic space environment, especially in terms of mechanical energy. This is in contrast to compact exoplanetary systems for which the planetary orbits likely lie inside the Alfvén surface, and thus in the subalfvénic wind regime. This could indicate that the radiometric scaling law might possibly only hold in superalfvénic conditions, and will be a topic of further studies.

\subsection{Effects of model assumptions}
After the LOFAR radio detections~\citep{2021A&A...645A..59T},  there has been multiple non-detections reported~\citep{2023arXiv231005363T,2024A&A...688A..66T,2025arXiv250106301C}.  Several reasons for the non-detections have been proposed, including the magnetic cycles of \(\tau\)~Boötis~A, unfavourable viewing geometry and beaming conditions, and variations in the magnetosphere of \(\tau\)~Boötis~Ab.

In the literature, estimates have been made for the emissions from
\(\tau\)~Boötis~Ab:
\citet{2007ApJ...668.1182L} suggests an upper limit on the emitted flux from \qtyrange{135}{300}{\milli\jansky}, and
conducted a multi-epoch search that did not detect \(\tau\)~Boötis~Ab at
\qty{74}{\mega\hertz}.
 \citet{1999JGR...10414025F} suggested that the median flux density should be \qty{\sim 2.2}{\milli\jansky} around \qty{28}{\mega\hertz} and pointed to how transient variations in wind speed and non-isotropic emissions could increase the flux by a factor of 100--1000. Similarly,  \citet{2005A&A...437..717G} suggested that the system should emit with a flux density of \qtyrange{4}{9}{\milli\jansky} and a frequency \qtyrange{7}{19}{\mega\hertz}.
\citet{2012MNRAS.423.3285V} applied a polytropic wind model driven by magnetograms from~\citet{2007MNRAS.374L..42C,2008MNRAS.385.1179D,2009MNRAS.398.1383F}, finding radio flux density values from \qtyrange{0.5}{1}{\milli\jansky}.
The recent study by \citet{2023A&A...671A.133E} applied a polytropic wind model~\citep{2016MNRAS.459.1907N} to obtain a radio flux density \(\phi  = \qty{300}{\milli\jansky}\)
only compatible with the observations of \citet{2021A&A...645A..59T} with a magnetospheric Poynting flux-to-auroral radio efficiency $\epsilonreplacement_\Aurora > 10^{-3}$ (ten times the value used in this work) or a stellar wind density orders of magnitude larger than the density of the predicted wind \citep{2016MNRAS.459.1907N}.
This result suggests that magnetospheric emissions from such close-in exoplanets are barely detectable, despite the star-planet interaction being extremely energetic, on the order of \(\qty{e18}{\watt}\). Here we explore how different model assumptions would have modified our results.

\subsubsection{Best-case efficiency factors}\label{sec:efficiency-factors}
In this work we have consistently applied best-case efficiency factors, solid angles of emission, and signal bandwidths. The computed radio flux densities scale linearly or inversely with each of these factors.

The efficiency factors \(\eta\) directly affect the radio intensity values shown in Fig.~\ref{fig:boxplot-power}. In this work we have applied the \(\eta_\Kinetic=\num{1e-5}\) and \(\eta_\Magnetic=\num{3e-3}\) values of~\citet{2007P&SS...55..598Z,2018A&A...618A..84Z}.
The latter value is a good fit for observations of both the outer planets and the Jovian moons. The review by~\citet{2018haex.bookE..22Z}, uses \(\eta_\Magnetic=\num{2e-3}\), however. \citet{2016MNRAS.461.2353N} have suggested a range of \(\eta_\Magnetic\) values from \numrange{e-4}{2e-3}. A larger efficiency factor such as the $\eta=\num{1e-2}$ used by \citet{2018ApJ...854...72T} would affect the results reported here, increasing the radio intensities by a factor of \num{\sim 3}. We consider such high efficiencies to be unlikely due to the numerous steps involved in converting incident stellar wind energy to auroral radio emission, such as reconnection efficiency, electron acceleration, wave-particle interaction and the ECMI mechanism. We thus consider our choice of \(\eta_\Magnetic=\num{3e-3}\) to be optimistic, but not unrealistically high.

For the resolved magnetospheric model, this work makes use of the best-case value for the  magnetospheric Poynting flux-to-auroral radio efficiency $\epsilonreplacement_\Aurora=\num{e-4}$~\citep{2023A&A...671A.133E} in the scope of the planetary magnetohydrodynamic model (Sect. \ref{sec:planetary-mhd}). From Jovian radio observations we have evidence that $\epsilonreplacement_\Aurora<\num{1e-4}$ \citep[see][]{2021A&A...655A..75S}; hence we consider our choice of $\epsilonreplacement_\Aurora=\num{e-4}$ optimistic.

\subsubsection{Solid angle of emission and signal bandwidth}\label{sec:solid-angle-and-bandwidth}
We have a applied a solid angle of emission \(\Omega=\qty{1.6}{\steradian}\)~\citep[see][]{2004JGRA..109.9S15Z} in all our calculations. In the literature, one accounts for the anisotropy\label{sec:anisotropy} of ECMI emissions by considering emissions in a thin, hollow cone pattern~\citep{2004JGRA..109.9S15Z}, so that the solid angle spanned by the (single) hollow cone shape is given by
\begin{equation}\label{eq:double-hollow-cone}
\Omega = 4\pi\sin\alpha\sin\delta/2,
\end{equation}
where \(\alpha\) is the cone's half-opening angle and \(\delta\) is its thickness. Based on solar system considerations,~\citet{2004JGRA..109.9S15Z} used an opening angle \(\alpha\) in the range \qtyrange{60}{90}{\degree} and a thickness \(\delta = \qty{17.5}{\degree}\). Applying these values gives \(\Omega\) ranging from \qtyrange{1.6}{1.9}{\steradian}, i.e., the signal is emitted in the direction of \qtyrange{13}{15}{\percent} of the sky.
A double hollow cone (one hollow cone for each planetary magnetic pole) of solid angle \(2\Omega\) is sometimes considered~\citep[as in e.g.][]{2019MNRAS.485.4529K}.

We note that in a spatially resolved model of the ECMI emission, such as that of \citet{2008GeoRL..3513107H,2010P&SS...58.1188H} we would apply a spatial integral of thin \(\delta\sim\qty{1}{\degree}\), local cones to model the emission pattern as in \citet{2004P&SS...52.1455Z} and recent work~\citep{2017ApJ...846...75P,2017GeoRL..44.9225L,2019A&A...627A..30L,2025arXiv250116180Z}. For planetary auroral emissions these local cones have differing orientations, and we apply \(\Omega=\qty{1.6}{\steradian}\), to be the union of all the local cones' emission directions on the celestial sphere, as in~\citet{2004JGRA..109.9S15Z,2019MNRAS.485.4529K}.

We have used the bursty signal bandwidth of \(\Delta f=\qty{6}{\mega\hertz}\) when computing flux intensities. A more conservative approach would use \(\Delta f\sim \max{f}\), (i.e. assuming that the signal is spread across a frequency band as wide as the maximum signal frequency). In our case this would reduce the calculated radio flux densities by a factor of \(21/6=3.5\). Overall, we see that by choosing conservative values of \(\Omega\) and \(\Delta f\) we would compute radio flux intensities \num{\sim 7} times lower than the ones shown in Fig.~\ref{fig:boxplot-power}.

We note, finally, that the \(\Omega=\qty{1.6}{\steradian}\) used by~\citet{2004JGRA..109.9S15Z} was computed based on a frequency range from \qtyrange{3}{16}{\mega\hertz}. The bursty signal of \citet{2021A&A...645A..59T} was detected in a narrower frequency range from \qtyrange{15}{21}{\mega\hertz}. If the emission cone thickness \(\delta\) decreases with decreasing frequency range, as suggested by~\citet{2004JGRA..109.9S15Z}, then \(\Omega\) would be smaller and the source intensity \(I\) values and the radio flux densities \(\phi\) would be greater than reported here by a factor \num{\sim2} (assuming a reciprocal relationship between \(\Delta f\) and \(\Omega\)).

\subsubsection{Magnetic cycles, magnetic mapping, and planetary phase}
The star \(\tau\)~Boötis~A is known for its magnetic variability and  cycles~\citep{2016MNRAS.459.4325M,2018MNRAS.479.5266J}; it is thus important to observe the stellar magnetic field in order to make predictions about a specific epoch such as those of the~\citet{2021A&A...645A..59T} observations. From Fig.~\ref{fig:intensity-sphere} we see that a radio flux intensity variation of a factor of \qty{\sim 5} can be expected depending on the magnetic map details, despite the well-characterised orbit and ephemeris of \(\tau\)~Boötis~Ab.
This variation demonstrates the importance of applying a contemporaneous magnetic map. By knowing the subplanetary point on the star we see that \(\tau\)~Boötis~Ab is not close to the astrospheric current sheet, which can be seen as a region of reduced magnetic radio intensity (and increased kinetic radio intensity). This information would not be available without the contemporaneous magnetic map presented in this work. Additionally it should be kept in mind that the average surface magnetic field strength, which is also found with Zeeman-Doppler imaging, induces a further flux intensity variation of \qty{\sim2} or more.

\section{Conclusions}\label{sec:conclusion}

In this work we have presented a three-dimensional magnetohydrodynamical wind model based on spectropolarimetric observations made at the same time as the radio detections in the \(\tau\)~Boötis system.

To account for transient variation in wind power and uncertainty surrounding the absolute magnetic field strength of \(\tau\)~Boötis~A, we have created four different stellar wind models. The B1 model uses the stellar magnetogram without any scaling applied. The next two models (B10 and B100) have their magnetic field scaled by a factor of 10 and 100, respectively. An additional model, SA10, where the Alfvén flux-to-field ratio is scaled by a factor of 10, is also presented. The wind models extend from the stellar chromosphere past the orbit of \(\tau\)~Boötis~Ab. We find that the planet orbits in the transalfvénic regime in the B1 model, in the subalfvénic regime for B10 and B100, and in the superalfvénic regime for the SA10 model.

For each of our model cases we trace the Alfvén characteristics backwards and forwards from the planetary magnetosphere, revealing different energy paths for the flow of Alfvén wave energy in the subalfvénic and superalfvénic regimes. In magnetic star-planet interactions, the radio emissions are not necessarily expected to originate from the planet itself, but may also originate from the stellar chromosphere at the footprints of the Alfvén characteristics~\citep[i.e. planet-induced emission, see the review by][]{2025ARA&A..63..299V}. These energy paths can thus be used for modelling magnetic star-planet interactions and their observable signatures in future work.

We have computed the expected radio emissions from \(\tau\)~Boötis~Ab based on the electron cyclotron maser instability (ECMI). For the radio-kinetic Bode's law-based mechanism we find that the emission strength scales sublinearly with increasing magnetic field strength, and that the signal is \num{\sim100} times too weak to reproduce the observed signal even under best-case conditions.
For the radio-magnetic Bode's law-based mechanism, we find that the emission intensity scales slightly superlinearly with the stellar magnetic field strength, and is not strongly affected by the Poynting flux-to-field ratio which controls the flux of Alfvén wave energy into the chromosphere. We find that a scaling of the magnetic field by a factor of \num{\sim10} is required to match the flux intensity range of the bursty signal reported by~\citet{2021A&A...645A..59T}. The radio-magnetic mechanism is thus the only out of the radio-kinetic and radio-magnetic mechanisms that is able to reproduce the bursty signal, albeit with a scaling of the stellar magnetic field \num{\sim10} and best-case conditions.

By modelling the stellar wind-planet interaction using a resolved planet-centred magnetohydrodynamic model, we assess the efficiency of energy transfer from the stellar wind towards the auroral Poynting fluxes. This efficiency \(T_\Aurora\) decreased from \num{2e-1} (B1) to about \num{5e-3} (B100), showing a drastic decrease for subalfvénic stellar wind conditions (models B1 and SA1). Assuming an auroral radio efficiency of $\eta^\prime_\Aurora = 10^{-4}$ (eq. \ref{eq:three_intensities}) we find the radio fluxes to be on the order of \qtyrange{0.7}{1.9}{\milli\jansky} with the flux increasing from the B1 model towards the B100 model. The estimated auroral radio fluxes more closely follow the radio-kinetic flux obtained from the radiometric Bode's law (Fig.~\ref{fig:boxplot-power}). This suggests that the auroral power input correlates mostly with the kinetic energy of the stellar wind, and that the radio-kinetic mechanism may still play a role in subalfvénic scenarios of our models. The resolved planetary model furthermore revealed that radio emission most likely originated from regions with planetary field lines pointing downstream due to reduced plasma density in the ionosphere. At the upstream-oriented field lines, stellar wind plasma is accumulated that would reflect radio emission originating from the ionosphere due to high plasma frequencies.

We conclude that the steady-state stellar wind presented here does not carry enough energy to power observable radio emission from \(\tau\)~Boötis~Ab in the range reported by \citet{2021A&A...645A..59T} unless a scaling \({\gtrsim} 10\) is applied to the surface magnetic field strength of the contemporaneous magnetic map that we have presented. This applies even in the best-case conditions.

Although the radio emission in the model presented here is able to escape the magnetosphere (see Fig.~\ref{fig:TauBooMagnetosphere}), an enhanced stellar wind density might result in enhanced magnetospheric plasma density which in turn may prohibit the escape of radio emission. Therefore, by increasing the energy density of the stellar wind and with it the plasma density, the resultant planetary radio emission might be increasingly shielded off by a densely populated magnetosphere.

It remains intriguing to see the bursty LOFAR signal occurring around quadrature (see Fig.~\ref{fig:phases}), as this configuration has been identified as the most favourable~\citep{2023MNRAS.524.6267K}. Given the magnetic cycles of \(\tau\)~Boötis~Ab~\citep{2016MNRAS.459.4325M,2018MNRAS.479.5266J}, and the suggested 100--1000 times flux intensity variations resulting from non-steady winds~\citep{1999JGR...10414025F}, it may be that favourable viewing geometry, favourable magnetic geometry, and a temporary surge of wind power are all simultaneously required to attain the flux levels reported by~\citet{2021A&A...645A..59T}. This may be the reason why follow-up observations of the \(\tau\)~Boötis system have not yielded positive detections and highlights the need for further concurrent radio observations and spectropolarimetric observations.

\section*{Acknowledgements}
This project has received funding from the European Research Council (ERC) under the European Union's Horizon 2020 research and innovation programme (grant agreement No 817540, ASTROFLOW and No. 884711, Exo-Oceans). This publication is part of the projects
OCENW.M.22.215 (under research programme Open Competition Domain Science - M) and VI.C.232.041 (under research programme Talent Programme Vici), which are financed by the Dutch Research Council (NWO).
We thank \href{www.surf.nl}{SURF (www.surf.nl)} for the support in using the National Supercomputer Snellius.
This work was carried out using the SWMF and BATS-R-US tools developed at the University of Michigan’s \href{https://clasp.engin.umich.edu/research/theory-computational-methods/center-for-space-environment-modeling/}{Center for Space Environment Modeling (CSEM)}. The modelling tools are available through the University of Michigan for download under a user license; an open source version is available at
\url{https://github.com/SWMFsoftware}.
In this work the freely available magnetohydrodynamic code PLUTO was employed, which is available at \url{https://plutocode.ph.unito.it/}.
This research has made use of NASA's \href{https://ui.adsabs.harvard.edu/}{Astrophysics Data System}.
This work has made use of the following additional numerical software, statistics software and visualisation software:
        NumPy~\citep{2011CSE....13b..22V},
        SciPy~\citep{2020SciPy-NMeth},
    Matplotlib~\citep{2007CSE.....9...90H},
\href{https://www.tecplot.com}{Tecplot}, and
\href{https://pypi.org/project/pytecplot/}{PyTecplot}.
We would like to express our gratitude to the anonymous reviewers for their valuable contributions, which have enhanced the quality and rigour of our article.

\section*{Data availability}
The data underlying this article will be shared on reasonable request to the corresponding author.

\let\mnrasl=\mnras
\bibliography{bibliography}

\appendix

\section{Spectropolarimetric observations}
Table~\ref{tab:obs} lists the spectropolarimetric observations used to reconstruct the stellar magnetic map. The times of the observations are given in coordinated universal time (UTC) and Julian date (JD). The cycle phase and stellar longitude of each observation is computed using the reference Julian date \num{2453450.984} and the stellar rotation period \num{3.31}.
\begin{table}
    \centering
    \caption{Time and phase for the spectropolarimetric observations used to construct the stellar magnetic map. The times of the observations are given in coordinated universal time (UTC) and Julian date (JD). The cycle phase and stellar longitude of each observations is computed using the reference Julian date \num{2453450.984}.}\label{tab:obs}
    \begin{tabular}{
        @{}
        l
        S[table-format=7.5]
        S[table-format=1.4]
        S[table-format=3.1]
        @{}
        }
    \toprule
    {UTC}& {Julian date} &  {Cycle Phase} & {Longitude} \\
    \midrule
    2017-02-15 05:06 & 2457799.71300 &  0.8154 & 66.5  \\
    2017-02-16 05:05 & 2457800.71217 &  0.1173 & 317.8  \\
    2017-02-17 03:30 & 2457801.64594 &  0.3994 & 216.2  \\
    2017-02-18 03:22 & 2457802.64060 &  0.6999 & 108.0  \\
    2017-02-19 03:19 & 2457803.63836 &  0.0013 & 359.5  \\
    2017-02-20 03:53 & 2457804.66188 &  0.3105 & 248.2  \\
    2017-02-22 03:49 & 2457806.65954 &  0.9141 & 30.9  \\
    \bottomrule
    \end{tabular}
\end{table}

\begin{figure}
    \centering    \includegraphics[width=\columnwidth]{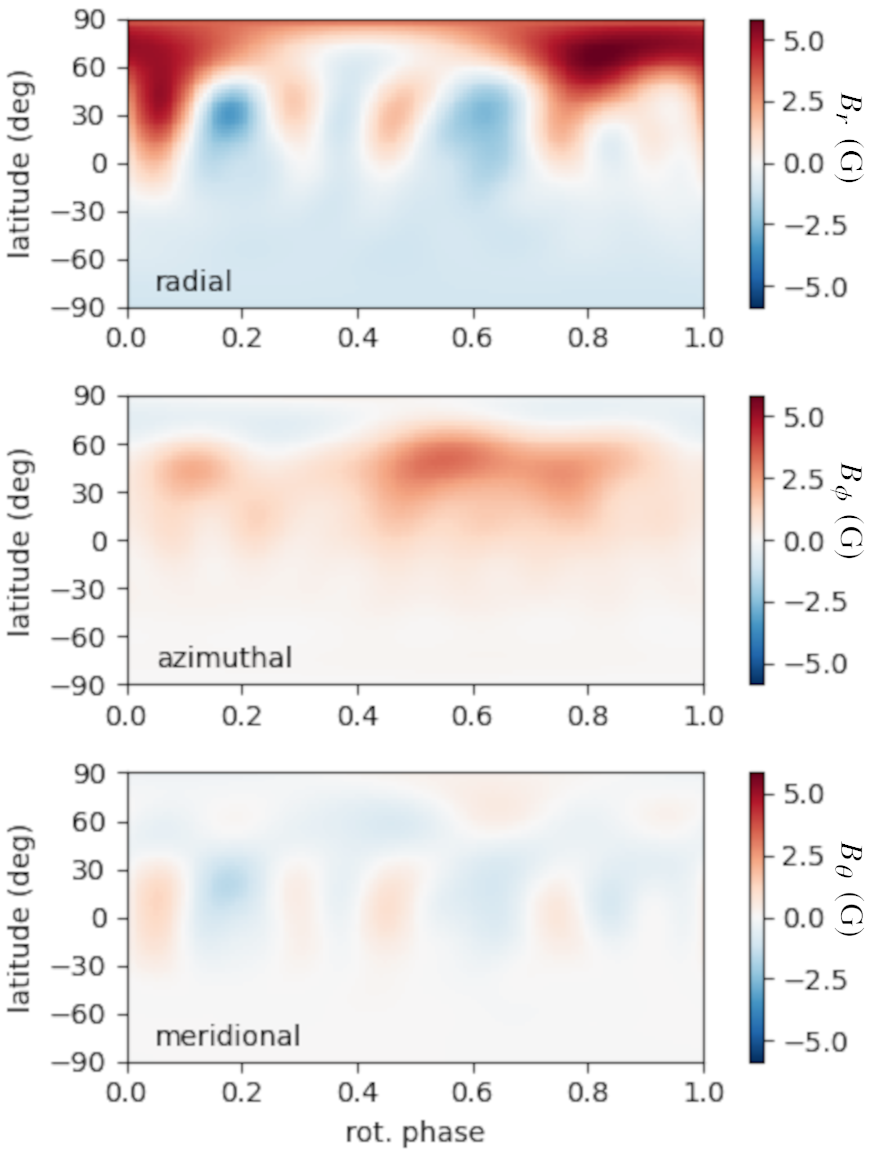}
    \caption{
        Radial, azimuthal, and meridional magnetic field strength reconstructed from the spectropolarimetric observations. The radial magnetic field strength in Fig.~\ref{fig:map} is taken from this vector magnetogram. Note that stellar longitude decreases with increasing observational phase in our model setup, thus the features appear horizontally flipped in comparison with Fig.~\ref{fig:map}. This plot was created using ZDIPy~\citep{2016MNRAS.457..580F,2018MNRAS.474.4956F}.
    }\label{fig:vector-magnetogram}
\end{figure}

\label{sec:vector-magnetogram}
For epoch~2017.12, the large-scale surface field had a mean strength of
\SI{1.6}{\gauss} and reached a maximum of \SI{5.9}{\gauss}.
The magnetic energy budget was dominated by the poloidal component
(\SI{74}{\percent} of the total), with the remaining
\SI{26}{\percent} stored in toroidal modes.
Within the poloidal field, the energy split into
\SI{40}{\percent} dipolar ($\ell=1$),
\SI{17}{\percent} quadrupolar ($\ell=2$),
\SI{14}{\percent} octupolar ($\ell=3$) and
\SI{30}{\percent} in higher-order terms ($\ell>3$).
The field was largely axisymmetric: modes with azimuthal
order $m=0$ contained \SI{52}{\percent} of the total magnetic energy,
comprising \SI{39}{\percent} of the poloidal and
\SI{88}{\percent} of the toroidal contributions.
Fig.~\ref{fig:vector-magnetogram} shows the radial, azimuthal, and meridional magnetic field strength reconstructed from the spectropolarimetric observations. This plot was created using ZDIPy\footnote{
ZDIPy is an open-source Python-based Zeeman-Doppler imaging code available on GitHub at \url{https://github.com/folsomcp/ZDIpy}.
}~\citep{2016MNRAS.457..580F,2018MNRAS.474.4956F}. The radial magnetic field strength in Fig.~\ref{fig:map} is taken from this vector magnetogram. Note that stellar longitude decreases with increasing observational phase in our model setup, thus the features appear left-right-flipped in comparison with Fig.~\ref{fig:map}.

\section{Average Alfvén radius}\label{sec:alfven-surface}

\begin{figure*}
    \centering
    \includegraphics{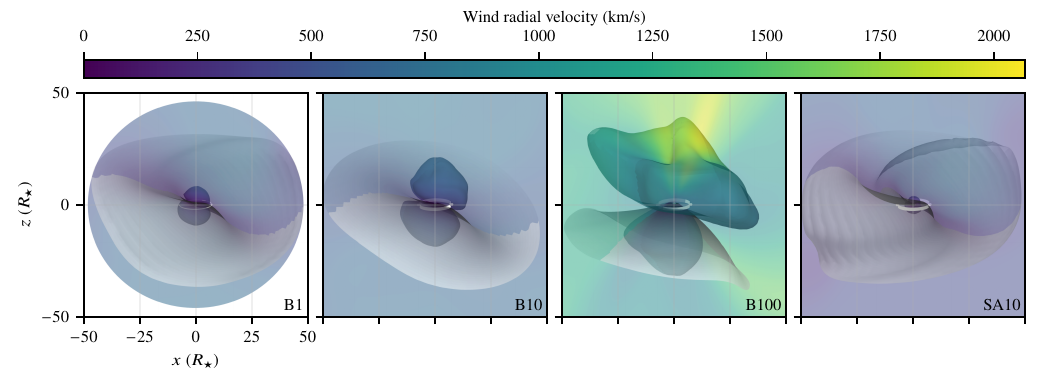}
    \caption{
        Wind radial velocity, Alfvén surface (coloured by radial velocity), and magnetic current sheet (grey) for the four model cases of Table~\ref{tab:models}. The planetary orbit is indicated as a white circle. The Alfvén surface increases with the magnetic field strength (from B1 to B100), and shrink with increased Poynting flux (cp. models B1 and SA10).
    }\label{fig:wind-ur}
\end{figure*}

Fig.~\ref{fig:wind-ur} shows the wind radial velocity, Alfvén surface, and magnetic current sheet for the four model cases of Table~\ref{tab:models}. The planetary orbit is indicated as a white circle.

The Alfvén surface is the set of points where the wind velocity \(u\) (in the stellar frame) is equal to the Alfvén velocity \(u_\Alfven = B/\sqrt{\mu_0 \rho}\).
In a low-\(\beta\) plasma
(see Section~\ref{sec:estimated-magnetosphere}) perturbations outside of the Alfvén surface cannot propagate starwards. The Alfvén surface can be seen to increase with the magnetic field strength, and shrink with increased Poynting flux.

The average Alfvén radius, reported in Table~\ref{tab:wind-aggregate-results}, is a measure of the average distance to the Alfvén surface. This value is computed by averaging the radial distance to the Alfvén surface over the stellar surface in the natural way as in~\citet{2022MNRAS.510.5226E}.

\section{Thermal bremsstrahlung and gyroemission}\label{sec:brems-and-gyro}
In this section we consider radio emissions due to thermal bremsstrahlung and gyroemissions.
\subsection{Thermal bremsstrahlung}
\begin{figure*}
    \centering
    \includegraphics{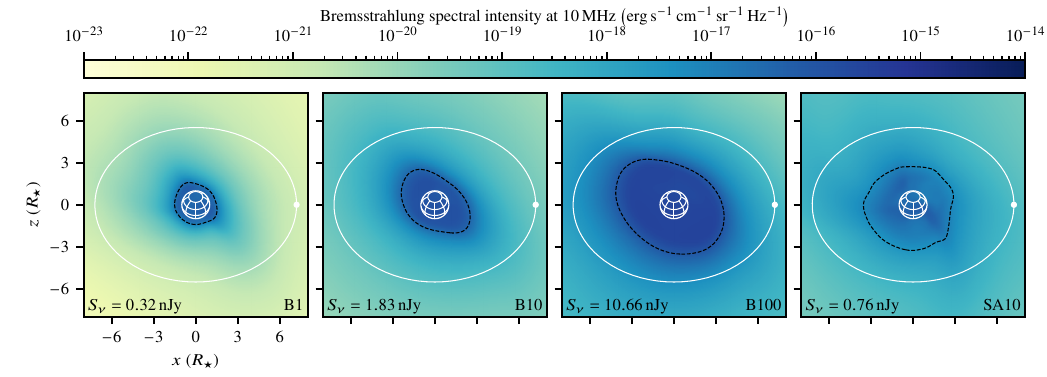}
    \includegraphics{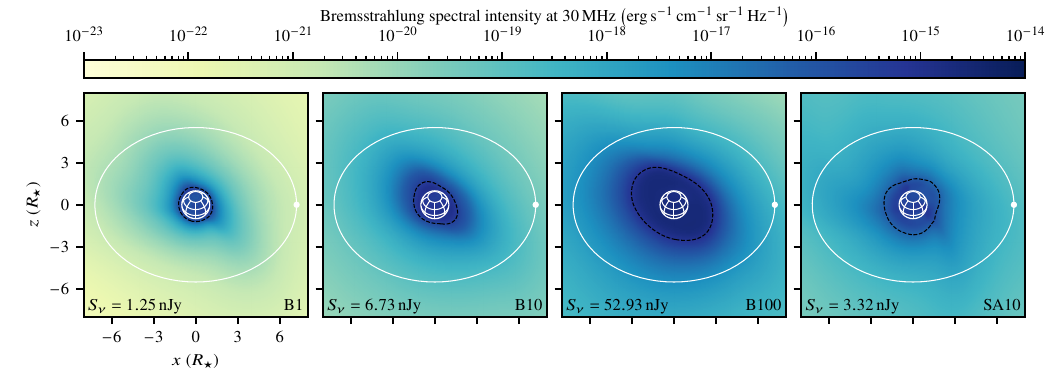}
    \caption{
        Thermal bremsstrahlung at \qty{10}{\mega\hertz} and \qty{30}{\mega\hertz}.
        The white graticule indicates the star's position, orientation, and size.
        The black dashed contour indicates the region inside of which the wind plasma is optically thick.
        When comparing the two frequencies, we see that the spectral intensity is lower at \qty{10}{\mega\hertz} than at \qty{30}{\mega\hertz}. The optically thick region, however, is larger at \qty{10}{\mega\hertz} than at \qty{30}{\mega\hertz}.
        Notably, the planet is never inside or behind the region where the wind plasma is optically thick at any of the two wavelengths.
    }\label{fig:wind-bremsstrahlung}
\end{figure*}
The stellar wind emits free-free radiation (thermal bremsstrahlung) at radio frequencies.
For thermal bremsstrahlung emission the emissivity is often simplified to~\citep[e.g.\ ][]{1986rpa..book.....R}
\begin{equation*}
    \varepsilon_\nu \propto n_\text{e}^2 T_\text{e}^{-1/2} \exp{\left(h\nu \middle/ k_\textsc{b} T_\text{e},\right)}
\end{equation*}
where \(\nu\) is the emission frequency, \(n_\text{e}\) is the electron density, \(T_\text{e}\) is the electron temperature. In Fig.~\ref{fig:wind-bremsstrahlung} we show the thermal bremsstrahlung emission at \qty{10}{\mega\hertz} for the wind models of Table~\ref{tab:models}, computed using the Radiowinds\footnote{
    The Radiowinds code may be found on GitHub at \url{https://github.com/ofionnad/radiowinds}.
} code~\citep{2019MNRAS.483..873O,2021ascl.soft01004O}.

In Fig.~\ref{fig:wind-bremsstrahlung} the black dashed contour indicates the region where the wind plasma is optically thick; we see that \(\tau\)~Boötis~Ab never enters the optically thick region at \qtyrange{10}{30}{\mega\hertz} and thus cannot be said to eclipse the star at \qtyrange{10}{30}{\mega\hertz}. Similarly the planet is not eclipsed by the star due to the system's rotational inclination (\qty{40}{\degree}, see Table~\ref{tab:properties}).
Notably, while emission strength is weaker at \qty{10}{\mega\hertz} than at \qty{30}{\mega\hertz}, the optically thick region is larger at \qty{10}{\mega\hertz} than at \qty{30}{\mega\hertz}.

Steady-state thermal bremsstrahlung emissions and variations therein induced by transiting planets have been studied by \citet{2018AJ....156..202C,2018ApJ...867...51M,2019MNRAS.483..873O}. Transiting planets are believed to induce variations in bremsstrahlung radio emissions, which \citet{2018ApJ...867...51M} found could reach up to unity for certain frequency ranges. The signals strengths of the thermal bremsstrahlung at \qty{10}{\mega\hertz} and \qty{30}{\mega\hertz} are given in Table~\ref{tab:wind-aggregate-results}. The values, of the order of \unit{\nano\jansky}, are however very low compared with the signals observed. This is in agreement with, e.g.,~\citet{2019MNRAS.484..648P} who considers exoplanet radio transits to be too weak to be observed even with the next generation of radio telescopes.

\begin{figure}
    \centering
    \includegraphics{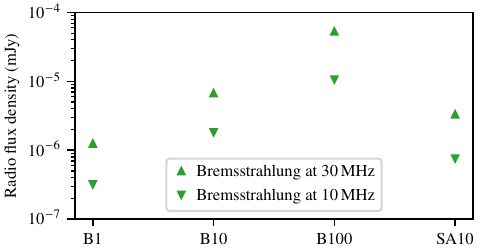}
    \caption{
        Radio flux densities from thermal bremsstrahlung.
        The intensities are about \num{e-6} times to weak to match the observed slowly varying signal
        L570725 of \citet{2021A&A...645A..59T}.
    }\label{fig:brems-power}
\end{figure}

\subsection{Gyroemissions}
Gyroemissions are caused by the gyration of electrons at their thermal velocities, and is a separate emission mechanism from thermal bremsstrahlung. Gyroemissions at the Sun are primarily associated with active regions and flares, and can be used to gauge their magnetic field strength~\citep{1997SoPh..174...31W,2007SSRv..133...73L}.
At the Sun, gyroemissions are generated when hot \qty{e6}{\kelvin} plasma interacts with strong magnetic fields \qty{100}{\gauss} or more~\citep{2019AdSpR..63.1404N}.
In the case of a non-flaring Sun, however, the radio emissions come from thermal bremsstrahlung, which tends to dominate over gyroemissions at frequencies up to \qty{3}{\giga\hertz}~\citep{1997SoPh..174...31W}. Due to the low frequency \(\leq\qty{30}{\mega\hertz}\) signals considered in this work and dominance of thermal bremsstrahlung below \qty{3}{\giga\hertz} at the Sun, we consider the calculations of gyroemissions to be beyond the scope of this paper and refer the reader to the work of e.g.~\citet{2021ApJ...914...52F}.

\section{Dipolar planetary magnetic field}\label{sec:dipole}
Assuming that the planet's magnetic field is dipolar, the planetary magnetic field is given by
\begin{equation}\label{eq:mag-dipole-vector}
\vec B(\vec r) = \frac{\mu_0}{4\pi} \left( \frac{3 \vec r (\vec m \cdot \vec r)}{r^5}-\frac{\vec m}{r^3} \right)
\end{equation} where \(\vec m\) is the dipole magnetic moment of the planet. By defining \(\alpha\) as the angle between \(\vec m\) and \(\vec r\), \(m r \cos \alpha = \vec m \cdot \vec r\), and taking the magnitude of equation~\eqref{eq:mag-dipole-vector} we find that the the planetary magnetic field strength is
\begin{equation*}
    B(r, \alpha) =  \frac{\mu_0}{4\pi}  \frac{m}{r^3} \sqrt{1 + 3\cos^2 \alpha}.
\end{equation*}
We can relate \(m\) to the  magnetic field strength at the planet's surface.
At the planet's magnetic pole \((R_\Planet, \qty{0}{\degree})\) we define \(B_\Planet = B(R_\Planet, \qty{0}{\degree})\). This sets the magnitude of the magnetic moment
\begin{equation*}
    m = \big({2\pi}\big/{\mu_0}\big) B_\Planet R_\Planet^3,
\end{equation*} where \(R_\Planet\) is the planet's radius and \(r\) is the distance from the planet's centre.
With the magnetic moment \(m\) thus expressed, we can express the magnetic field strength at any point \((r, \alpha)\) as a function of \(B_\Planet\):
\begin{equation}\label{eq:mag-dipole-strength}
    B(r, \alpha) = \frac{B_\Planet}2  \left(\frac{R_\Planet^3}{r}\right)\sqrt{1 + 3\cos^2 \alpha}.
\end{equation}
We see from the square root term that \(B(r, \alpha)\) varies by a factor of 2 between the magnetic pole, where \(\alpha=\qty{0}{\degree}\) and the magnetic equator, where \(\alpha=\qty{90}{\degree}\) for all distances \(r\).

\subsection{Magnetospheric size}\label{sec:magnetosphere-size}
The planet-produced pressure is calculated analytically based on a number of simplifying assumptions. We assume that the planet-produced pressure is dominated by the magnetic pressure. From eq.~\eqref{eq:mag-dipole-strength} we see that the magnetic field over the surface scales as
\(B(r, \alpha) = (R_\Planet/r)^3 B(R_\Planet, \alpha) \). The planet-produced pressure is then given by
\begin{equation*}
    p_\Planet(r, \alpha)= \frac{1}{2\mu_0} B(r, \alpha)^2
    =
    \frac{1}{2\mu_0}
    \frac{B_\Planet^2}{4}
    \left( \frac{R_\Planet}{r} \right)^6
    \left(1+3\cos^2\right).
\end{equation*}
Equating the wind pressure \(p_\Wind\) to the planet-produced pressure \(p_\Planet\) and solving for \(r\) gives an indication of the size of the planetary magnetosphere \(R_\Mag\),
\begin{equation*}
    \frac{R_\Mag}{R_\Planet} = \left( \frac{1}{2\mu_0} \frac{B_\Planet^2}{p_\Wind} \right)^{1/6} \left(\frac{1+3\cos^2\alpha}{4}\right)^{1/6}
    \simeq
    \left( \frac{1}{2\mu_0}\frac{B_\Planet^2}{p_\Wind} \right)^{1/6}.
\end{equation*}
The geometric term containing \(\alpha\) varies from \numrange{0.79}{0.89}. We note that for superalfvénic flow a term \(\xi\) is often included to account for the effects of electrical currents in the magnetopause~\citep{2004pssp.book.....C}. The term \(\xi = 2^{1/3}\) is used in~\citet{2011AN....332.1055V}.

\subsection{The size of the polar cap/auroral region}\label{sec:polar-cap}

The auroral region is a ring-shaped region around the magnetic poles of the planet. The outer boundary of the auroral ring is thought to be the boundary between open and closed planetary magnetic field lines, i.e. magnetic field lines with one (open) or both (closed) ends connected to the planet~\citep{2001JGR...106.8101H}. This region is also referred to as the polar cap~\citep{2010Sci...327.1238T,2013A&A...557A..67V}.

Since the dipolar magnetic field of the planet is disturbed by the stellar wind beyond the magnetopause distance \(R_\Mag\) we can trace the dipolar planetary magnetic field lines to see whether they extend past \(R_\Mag\), in which case they may connect with the magnetic field embedded in the stellar wind. Given that the distance to the magnetopause is \(R_\Mag\), the polar cap is the region where the magnetic field lines extend past \(R_\Mag\).

We consider the dipolar magnetic field of eq.~\eqref{eq:mag-dipole-vector}. Using coordinates where \(\vec m \parallel \uvec z\) we have
\begin{equation*}
B_x = \frac{\mu_0}{4\pi} \left(3 m x z\right)/ r^5
\quad \text{and} \quad
B_z =
\frac{\mu_0}{4\pi}(2 m z^2-x^2)/r^5.
\end{equation*}
The magnetic field lines in the \(xz\) plane are described by
\begin{equation*}
\frac{\mathrm{d}z}{\mathrm{d}x} = \frac{B_z}{B_x} = \frac{2z^2-x^2}{3xz},
\end{equation*}
which is a Bernoulli differential equation with a family of solutions parametrised by \(r_0\):
\begin{equation*}
    z = \pm \sqrt{r_0^{2/3} x^{4/3} - x^2},
\end{equation*}
with each value of \(r_0\) giving rise to a different field line. Along each field line in the \(xz\) plane \(r^2 = x^2 + z^2 = r_0^{2/3} x^{4/3}\), and thus
\begin{equation}\label{eq:dipole-fieldline-distance}
    r = r_0 \sin^2{\alpha}
\end{equation}
by cylindrical symmetry, where we used \(x = r \sin\alpha\) with \(\alpha\) being the colatitude or polar angle. We see that \(r_0\) is the maximum distance reached by the fieldline, reached when \(\alpha=\qty{90}{\degree}\), i.e. in the magnetic equatorial plane.

To find the colatitude of the polar cap \(\alpha_\text{cap}\) we find the \(\alpha\) value at \(r=R_\Planet\) for the field line that reaches \(r_0=R_\Mag\), i.e. where \(R_\Planet=R_\Mag \sin^2 \alpha_\text{cap}\) by eq.~\eqref{eq:dipole-fieldline-distance}. This gives
\begin{equation}\label{eq:polar-cap-colatitude}
    \alpha_\text{cap} = \arcsin{\sqrt{R_\Planet/R_\Mag}}
\end{equation}
as given in e.g.~\citet{2011AN....332.1055V}.

\bsp
\label{lastpage}

\end{document}